\documentclass[a4paper, 11pt]{article}

\usepackage{graphicx}
\usepackage{epsf,amsmath,bbold,amsfonts,stmaryrd}
\usepackage{multirow}
\usepackage{makecell} 
\usepackage[utf8]{inputenc}
\usepackage{mathrsfs}
\usepackage{appendix}
\usepackage{amssymb}
\usepackage{float}
\usepackage{xcolor}
\usepackage{cite}
\usepackage{hyperref}
\hypersetup{pageanchor=false}
\usepackage{indentfirst}
\usepackage{url}
\usepackage{float}
\usepackage{caption}
\usepackage[numbers,square,comma,sort&compress,merge]{natbib}
\usepackage{esint}
\usepackage{subfigure}

\usepackage{ulem}

\def\be{\begin{equation}}
\def\ee{\end{equation}}

\def\bea{\begin{eqnarray}}
\def\eea{\end{eqnarray}}

\def\ba{\begin{array}}
\def\ea{\end{array}}

\def\bc{\begin{center}}
\def\ec{\end{center}}

\def\bl{\begin{flushleft}}
\def\el{\end{flushleft}}

\def\br{\begin{flushright}}
\def\er{\end{flushright}}

\def\bi{\begin{itemize}}
\def\ei{\end{itemize}}

\def\bt{\begin{tabular}}
\def\et{\end{tabular}}

\newcommand{\df}{\mathrm{d}}

\numberwithin{equation}{section}

\begin{document}

\title{Critical Behavior of Photon Rings in Kerr-Bertotti-Robinson Spacetime}

\author{Xi Wan$^{1}$, Zhenyu Zhang$^{1\ast}$, Fang-Stars Wei$^{1}$, Yehui Hou$^{2}$, Bin Chen$^{1,3\dagger}$
}
\date{}

\maketitle

\vspace{-10mm}

\begin{center}
{\it
	$^1$ Institute of Fundamental Physics and Quantum Technology, and School of Physical Science and Technology,
	Ningbo University, Ningbo, Zhejiang 315211, P. R. China\\\vspace{4mm}

	$^2$ Tsung-Dao Lee Institute, Shanghai Jiao-Tong University, Shanghai 201210, P. R. China\\\vspace{4mm}

	$^3$ School of Physics, and Center for High Energy Physics, Peking University,
	No.5 Yiheyuan Rd, Beijing 100871, China\\\vspace{4mm}
}
\end{center}

\vspace{8mm}

\begin{abstract}

In this work, we investigate the critical behavior of photon rings in the Kerr–Bertotti–Robinson spacetime, describing a rotating black hole immersed in a background magnetic field. We analyze the radial and angular motions of photons under the small magnetic field approximation. Focusing on unstable spherical orbits, we determine three key parameters, $\gamma$, $\delta$, and $\tau$, which characterize radial compression, azimuthal advancement, and time delay. We then examine how these parameters depend on the black hole spin, magnetic field strength, and observer inclination for both on-axis and off-axis observers, and we further analyze the properties of higher-order images through near-critical lens equations. The results show that the magnetic field modifies the geodesic structure, and leads to observable changes in the fine structure of photon rings, providing a useful framework for probing magnetized black hole environments.

\end{abstract}

\vfill{\footnotesize $\ast$ Corresponding author: zhangzhenyu@nbu.edu.cn

\footnotesize $\dagger$ Corresponding author: chenbin1@nbu.edu.cn}
\maketitle

\newpage
\baselineskip 18pt

\section{Introduction}\label{sec1}

Magnetars demonstrate that ultra-strong magnetic fields, reaching $10^{14}-10^{15}$ Gauss, can exist in astrophysical environments \cite{Olausen:2013bpa,Kaspi:2017fwg}. In this context, it is plausible that similarly strong fields may arise near black holes. Observationally, the magnetar SGR J1745–29, located within $\sim 0.1$ pc of SgrA*, shows that highly magnetized objects can exist in close proximity to a supermassive black hole \cite{Eatough:2013nva,Kennea:2013dfa}. Theoretically, processes such as core collapse and compact object mergers can generate or amplify magnetic fields to comparable strengths \cite{Duncan:1992hi,Price:2006fi,Mosta:2015ucs,Kiuchi:2015qua, Ruiz:2016rai}. Such fields can significantly influence accretion, particle dynamics, and energy extraction, motivating the study of black holes immersed in strong magnetic fields.

When the backreaction of the magnetic field on the spacetime geometry is negligible, black hole magnetospheres can be described by vacuum solutions, including Wald's uniform magnetic field  \cite{Wald:1974np}, configurations with more general field orientations \cite{bivcak1976stationary, bivcak1977stationary, bivcak1980stationary, bivcak1985magnetic}, and force-free electrodynamic solutions \cite{Blandford:1977ds, Brennan:2013kea, Gralla:2014yja}. 
In the strong-field regime, where the gravitational effect of the magnetic field becomes non-negligible, exact magnetized solutions are required, most notably the Ernst solutions \cite{Ernst:1976mzr} and the Kerr–Melvin spacetime \cite{Ernst:1976bsr}, a prototypical model of a rotating black hole immersed in a magnetic universe. 
However, their nontrivial asymptotic structure and algebraic complexity hinder detailed analyses of particle dynamics and associated observable signatures \cite{Stuchlik:1999mro,Li:2018wtz,Wang:2021ara,Junior:2021dyw,Hou:2022eev}.

More recently, a new exact solution, known as the Kerr-Bertotti-Robinson (KBR) spacetime \cite{Podolsky:2025tle,Ovcharenko:2025cpm}, has attracted considerable attention \cite{Wang:2025vsx,Vachher:2025jsq,Ali:2025beh,Zhang:2025ole,Wang:2025bjf,Liu:2025wwq,Li:2025rtf,Zeng:2025tji,Zeng:2025olq,Mirkhaydarov:2026fyn,Wang:2026czl,Hu:2026slp,Rehman:2026rzq,Xamidov:2026kqs,Lu:2026kcm}. This spacetime geometry describes a rotating black hole embedded in the Bertotti-Robinson universe, which represents an asymptotically uniform electromagnetic field. 
In contrast to the Kerr–Melvin case, the KBR spacetime belongs to the Petrov type D class \cite{Petrov:2000bs}, reflecting enhanced hidden symmetries \cite{Gray:2025lwy}. It also admits separability of the Hamilton-Jacobi equation for null geodesics \cite{Wang:2025vsx}, which greatly facilitates analyses of photon motion, unstable spherical photon orbits, and black hole shadow formation.

The strong gravitational lensing of black holes leaves distinctive imprints on black hole images, most notably photon rings, which have been widely studied following the Event Horizon Telescope observations \cite{Perlick:2021aok,Chang:2021ngy,Kuang:2022ojj,Yuan:2024wdl,Hu:2022lek,Guo:2022muy,Zhang:2022osx,Wang:2024uda,Zhang:2024jrw,Li:2026ogy,Guo:2024mij,Zhang:2024lsf}. It has also been recognized that such effects can influence spatially unresolved observables due to their intrinsically long photon paths, time delays, and frequency shifts \cite{Wong:2024gph, Palumbo:2025tyu}. The critical behavior of photons forming the photon ring is characterized by three parameters \cite{Gralla:2019drh,Hou:2022gge}: the Lyapunov exponent $\gamma$, which governs the exponential demagnification per half-polar oscillation \cite{Cardoso:2008bp,Johnson:2019ljv}; the rotation parameter $\delta$, encoding the spacetime-induced azimuthal shift between successive trajectories \cite{Teo:2003ltt}; and the time-delay parameter $\tau$, describing the temporal separation between successive loops. In future high-resolution observations \cite{Tiede:2022grp,Johnson:2024ttr,Lupsasca:2024xhq}, these quantities can be extracted via photon-ring autocorrelation analyses \cite{Hadar:2020fda,Chen:2022nbb,Zhang:2025vyx,Cardenas-Avendano:2024sgy}, providing direct probes of the underlying spacetime geometry.

Motivated by the prominent imprints of photon rings across diverse astrophysical black hole environments and forthcoming observational efforts, we present a detailed analysis of the critical photon-ring parameters in the KBR spacetime, with particular emphasis on their dependence on an external magnetic field. By exploiting the first-order null geodesic equations and their representation in terms of elliptic integrals, we derive analytic expressions for the parameters $\gamma$, $\delta$, $\tau$ through a systematic study of near-critical photon trajectories. We further obtain explicit perturbative expansions in the small-field limit. By examining the variation of $\gamma$, $\delta$ and $\tau$ with the magnetic parameter, we elucidate how strong magnetic fields influence both the underlying spacetime structure and the observable characteristics of photon rings in the strong-field regime.

The remaining part of this paper is organized as follows. In Sec.~\ref{sec:metric}, we first review the metric form and basic properties of the KBR spacetime and present the photon geodesic equations. Sec.~\ref{sec:geo} then provides a detailed analysis of the radial and angular motions of photons. In Sec.~\ref{sec:paras}, we derive general expressions for the three key parameters that characterize the photon ring. Sec.~\ref{sec:keyparares} further explores how these parameters vary with the black hole spin, magnetic field strength, and observer inclination, for both on-axis and off-axis observers. The properties of higher-order images are analyzed in Sec.~\ref{sec:lens} through the study of near-critical lens equations. Finally, we summarize our main results in Sec.~\ref{sec:sum}. Some technical details are presented in the appendices.

In this paper, we work in units where $G=c=1$. Under this convention, the relation between the magnetic field $B$ and its value in Gauss is given by:
\begin{equation}\label{Bgauss}
B_{\text{Gauss}} = \frac{c^4}{G^{3/2} M_\odot}\left( \frac{M_\odot}{M} \right) (BM)  = 2.36 \times 10^{19} \left( \frac{M_\odot}{M} \right) (BM).
\end{equation}
For a supermassive black hole with a mass on the order of $M \sim 10^9 M_\odot$, a dimensionless field strength of $BM = 0.01$ translates to a physical magnetic field of approximately $B_{\text{Gauss}} \sim 10^8 \ \text{Gauss}$. In addition, we set $M=1$ in the remaining part of this paper for simplicity.

\section{The KBR spacetime}\label{sec:metric}

In this section, we aim to provide a brief review of the KBR spacetime \cite{Podolsky:2025tle} and the null geodesic in it. The line element of the KBR black hole can be expressed as 
\bea
\begin{aligned}
	\df s^2 &=\dfrac{1}{\Omega^2} \Big[-\frac{Q}{\Sigma}
	\big(\df t - a \sin^2\theta\, \df\phi\big)^2
	+ \frac{\Sigma}{Q}\, \df r^2
	+ \frac{\Sigma}{P}\, \df\theta^2
	+ \frac{P}{\Sigma} \sin^2\theta\,
	\big(a\, \df t - (r^2 + a^2)\, \df\phi\big)^2 \,\Big] \\
 	&= g_{\mu\nu} \df x^\mu \df x^\nu.
\end{aligned} 
\eea
The functions in the metric are given by 
\bea
\begin{gathered}
	\Sigma(r,\theta)  = r^2 + a^2 \cos^2\theta\,, \quad
	P = 1 + B^2 \left(\dfrac{I_2}{I_1^2} - a^2 \right) \cos^2\theta\,,\quad
	Q(r) = \left(1 + B^2 r^2 \right) \Delta(r)\,, \\
	\Omega^2(r,\theta) = \left(1 + B^2 r^2 \right) - B^2 \Delta(r) \cos^2\theta\,, \quad
	\Delta(r) = \left(1 - B^2 \dfrac{I_2}{I_1^2}\right) r^2 - 2\dfrac{I_2}{I_1}\, r + a^2\,,
\end{gathered} 
\eea
with  
\begin{align}
	I_1 = 1 - \tfrac{1}{2} B^2 a^2\,, \qquad
	I_2 = 1 - B^2 a^2,
\end{align}
where $a$ represents its spin parameter, and $B$ is used to describe the strength of the electromagnetic field. When the magnetic field $B$ vanishes, it reverts to the Kerr spacetime. The event horizons are given by $g^{rr}=0$, corresponding to $\Delta=0$, which yields up to two solutions: 
\begin{equation}
	r_{\pm} = \frac{I_2 \pm \sqrt{I_2 - a^2 I_1^2 }}{I_1^2 - B^2 I_2}I_1\, ,
\end{equation}  
where $r_\pm$ correspond to the radii of outer and inner horizons, respectively. 

We then proceed to discuss photon geodesics near the KBR black hole. As stated in \cite{Wang:2025vsx}, the Hamilton–Jacobi equation for null geodesics in the KBR spacetime is separable, and therefore there exists an additional conserved quantity, namely the Carter constant $C_Q$. For photons, it is convenient to define the dimensionless conserved quantities as $\lambda = L/E$ and $\eta = C_Q/E^2$ where $E=-p_t$ and $L=p_\phi$. This allows us to express the geodesic equations as \cite{Wang:2025vsx}
\begin{equation}
	\begin{aligned}
		\frac{\Sigma}{\Omega^2 E} p^t &= \frac{r^2 + a^2}{Q(r)}\left(r^2 + a^2 - a\lambda\right) + \frac{a}{P}(\lambda - a \sin^2\theta), \\
		\frac{\Sigma}{\Omega^2 E}p^r &= \pm_r \sqrt{R(r)}, \\
		\frac{\Sigma}{\Omega^2 E} p^{\theta} &= \pm_\theta \sqrt{\Theta(\theta)}, \\
		\frac{\Sigma}{\Omega^2 E}p^{\phi} &= \frac{a}{Q(r)}\left(r^2 + a^2 - a\lambda\right) + \frac{\lambda}{P \sin^2\theta} - \frac{a}{P},
	\end{aligned}
\end{equation}
where 
\bea
\begin{aligned}
\mathcal{R}(r) &= \left[(a^2 + r^2) - a\lambda\right]^2 - \Delta\left[\eta + (\lambda - a)^2\right]- B^2 r^2 \Delta\left[\eta + (\lambda - a)^2\right],\label{rp}\\
\Theta(\theta) &= \eta - \left(\lambda^2 \csc^2\theta - a^2 \right)\cos^2\theta + B^2\left(\frac{I_2}{I_1^2} - a^2\right)\left[(\lambda - a)^2 + \eta\right] \cos^2\theta\,.
\end{aligned} 
\eea
By introducing the Mino time $\mathcal{T}$ \cite{Mino:2003yg}, namely, 
\begin{equation}
	\frac{dx^{\mu}}{d\mathcal{T}}=\frac{\Sigma}{\Omega^2 E} p^{\mu},
\end{equation}
the geodesic equations can be recast into the forms of integral
\bea
\begin{aligned}
	\mathcal{T} &=I_r=G_{\theta}\label{Mino},\\
	\Delta\phi &=\phi_o-\phi_s=I_{\phi}+G_{\phi}\label{dphi},\\
	\Delta t &=t_o-t_s=I_t+a G_t\label{deltat},
\end{aligned}
\eea
where the geodesic integrals are given by
\bea
\begin{gathered}
I_{r}  =\fint_{r_{s}}^{r_{o}} \frac{\df r}{\pm_r \sqrt{\mathcal{R}(r)}},\quad
G_{\theta} = \fint_{\theta_{s}}^{\theta_{o}} \frac{\df\theta}{\pm_\theta \sqrt{{\Theta}(\theta)}},\\
I_{\phi} =\fint_{r_{s}}^{r_{o}}\frac{a(r^2+a^2-a \lambda)}{\pm_r \sqrt{\mathcal{R}(r)} Q(r)} \df r \label{I_{phi}},\quad
G_{\phi} =\fint_{\theta_{s}}^{\theta_{o}} \left( \frac{\lambda}{P \sin^2\theta} - \frac{a}{P}\right)  \frac{d\theta}{\pm_\theta \sqrt{{\Theta}(\theta)}}, \\
I_t = \fint_{r_{s}}^{r_{o}} \frac{(r^2 + a^2)\left(r^2 + a^2 - a\lambda\right) }{{\pm_r \sqrt{\mathcal{R}(r)}} Q(r)} \df r,\quad
G_t =\fint_{\theta_{s}}^{\theta_{o}}\frac{ (\lambda - a \sin^2\theta)}{\pm_\theta \sqrt{{\Theta}(\theta)} P}  \df\theta.
\end{gathered}
\eea
The subscripts ``$s$'' and ``$o$'' represent the source and the observer, respectively, i.e., the starting and ending points of the geodesic integrals.

\section{Detailed analysis of the geodesic integrals}\label{sec:geo}
In order to compute the three key parameters later, we first need to carefully analyze the behavior of null geodesics.

\subsection{Angular integrals}\label{ai}

Introducing the variable $u = \cos^2{\theta}$, the angular potential can be recast as
\begin{equation}
	\Theta(u) = \frac{1}{1-u} \left(\eta + A_1 u - A_2 u^2\right) = \frac{A_2}{1-u}\left( u_+ - u\right) \left( u - u_- \right),
\end{equation}
where
\begin{equation}
\begin{aligned}
		u_{\pm} &= \frac{A_1 \pm \sqrt{A_1^2 + 4\eta A_2}}{2A_2},\\
		A_1 &= \left[a^2  - \lambda^2 - \eta - B^2 \left(a^2 - \frac{I_2}{I_1^2} \right)\left[(a - \lambda)^2  +\eta\right]\right]\,, \\
		A_2 &= \left[a^2 - B^2 \left(a^2 - \frac{I_2}{I_1^2} \right)\left[(a - \lambda)^2  +\eta\right] \right].
\end{aligned}\label{A}
\end{equation}
For small values of $B$ and with $a \neq 0$, we find $A_2 > 0$. When the magnetic field equals zero, it reverts to  Kerr black hole and only photons with $\lambda = 0$ can reach the polar axis. When $\eta < 0$, we have $u_- > 0$, confining photons to the intervals $\theta \in [\theta_- , \theta_+]$ or $ [\pi - \theta_+ , \pi - \theta_-]$ with $\theta_{\pm} = \arccos{\sqrt{u_{\pm}}}$. In contrast, for $\eta \geq 0$, we can obtain $u_- < 0$, so that $u \in [0, u_+]$, allowing photons to cross the equatorial plane ($u = 0$) and oscillate within $\theta \in [\pi -\theta_+ ,  \theta_+]$ with  $\theta_+ = \arccos{(-\sqrt{u_+})}$. 

In this work, we focus on the behavior near unstable spherical orbits. Therefore, only the case $\eta >0$ is relevant, which means the geodesic oscillates between the turning points that lies above and below the equator. The angular path integral can be expressed in terms of Eq.~\eqref{A} and the number of angular turning points $m$. 
Then the integrals can be expressed by
\begin{align}
	G_{\theta} &= \frac{1}{\sqrt{A_2}\sqrt{ -u_{-}}}\left[2mK \pm_{s} F_{s} \mp_{o} F_{o}\right],\label{a}\\
    G_{\phi} & = \frac{C_1}{\sqrt{A_2} \sqrt{-u_-}} \left[ 2m\Pi_1 \pm_s \Pi_{1s} \mp_o \Pi_{1o} \right] + \frac{C_2 - a}{\sqrt{A_2} \sqrt{-u_-}} \left[ 2m\Pi_2 \pm_s \Pi_{2s} \mp_o \Pi_{2o} \right] ,\label{b}\\
    G_t &=\frac{a}{C\sqrt{A_2}\sqrt{-u_-}} \left[ 2mK \pm_s F_s \mp_o F_o \right] + \frac{C(\lambda - a) - a}{C\sqrt{A_2}\sqrt{-u_-}} \left[ 2m\Pi_2 \pm_s \Pi_{2s} \mp_o \Pi_{2o} \right] \label{Gt},
\end{align}		
where
\bea
\begin{gathered}   
C_1 =\frac{\lambda}{1+C}, \quad C_2=\frac{\lambda C}{1+C}, \quad C= B^2 \left(\dfrac{I_2}{I_1^2} - a^2 \right).\\
F_{i} =F\left(\arcsin \left(\frac{\cos \theta_{i}}{\sqrt{u_{+}}}\right) \bigg\vert \frac{u_{+}}{u_{-}}\right),	\quad
K =K\left( \frac{u_{+}}{u_{-}}\right)=F\left( \frac{\pi}{2}\bigg\vert\frac{u_+}{u_-}\right) ,\\
\Pi_{1i}=\Pi\left(u_+;\arcsin \left(\frac{\cos \theta_{i}}{\sqrt{u_{+}}}\right) \bigg\vert \frac{u_{+}}{u_{-}}\right),\quad
\Pi_{2i}=\Pi\left(-C u_+;\arcsin \left(\frac{\cos \theta_{i}}{\sqrt{u_{+}}}\right) \bigg\vert \frac{u_{+}}{u_{-}}\right),\\
\Pi_1=\Pi\left(u_+\bigg\vert\frac{u_{+}}{u_{-}}\right)=\Pi\left(u_+;\frac{\pi}{2}\bigg\vert\frac{u_{+}}{u_{-}}\right),\quad
\Pi_2=\Pi\left(-Cu_+\bigg\vert\frac{u_{+}}{u_{-}}\right)=\Pi\left(-Cu_+;\frac{\pi}{2}\bigg\vert\frac{u_{+}}{u_{-}}\right).	
\end{gathered}
\eea 	
$F$ and $\Pi$ are elliptic functions of the first and third kinds, respectively.
The subscript $i$ in $\pm_i$ denotes the sign of $p^{\theta}$ evaluated at the source ($i=s$) and at the observer ($i=o$), which satisfy the relation $\pm_{s}=\pm_{o}(-1)^m$.

With the standard substitution $u=u_+\sin\psi$, the equatorial crossing condition $u=0$ corresponds to $\psi=0$, so the endpoint contributions $F_i$, $\Pi_{1i}$ and $\Pi_{2i}$ vanish. The polar turning point $u=u_+$ corresponds to $\psi=\frac{\pi}{2}$,  and the incomplete elliptic integrals can reduce to their complete forms $K$, $\Pi_1$ and $\Pi_2$ as
\begin{equation}
		F_i = \pm K, \quad \Pi_{1i} = \pm \Pi_1, \quad \Pi_{2i} =\pm \Pi_2.
\end{equation}
The orbital fraction $n$ can be defined as
\begin{equation}
	n=\frac{G_{\theta}}{2\int_{\theta_-}^{\theta_+}d\theta [\Theta(\theta)]^{-1/2}}=\frac{\sqrt{A_2}\sqrt{ -u_{-}}}{4 K} I_r\label{n}.
\end{equation}
Using Eq.~\eqref{Mino},~\eqref{a} and~\eqref{n}, we can establish the relationship between the number of angular turning points and the orbital fraction as
\begin{equation}
	n=\frac{1}{4K}\left[ 2m K \pm_{s}F_s\mp_oF_o\right] =\frac{1}{2}m\pm_{o}\frac{1}{4}\left[(-1)^m\frac{F_s}{K}- \frac{F_o}{K}\right].\label{n1}
\end{equation}

\subsection{Radial roots}
To study the characteristics of radial motion, it is necessary to analyze the roots of the radial potential $R(r)$, which is a quartic function
\begin{equation}
	\mathcal{R}(r)=\mathcal{A} r^4+\mathcal{B} r^3+\mathcal{C}r^2+\mathcal{D}r+\mathcal{G}.
\end{equation}
The coefficients in the potential are given by
\bea
\begin{aligned}
	\mathcal{A} &=1 - B^2 \left[1 + \frac{4 B^2 \left(-1 + a^2 B^2\right)}{\left(-2 + a^2 B^2\right)^2}\right] \left[\eta + ( \lambda-a)^2\right],\\
	\mathcal{B} &=\frac{4 B^2 \left(-1 + a^2 B^2\right) \left[(\lambda-a)^2 + \eta\right]}{-2 + a^2 B^2},\\
	\mathcal{C} &=2 a^2 - a^2 B^2 \left[\eta + ( \lambda-a)^2\right] + \left[-1 - \frac{4 B^2 \left(-1 + a^2 B^2\right)}{\left(-2 + a^2 B^2\right)^2}\right] \left[\eta + ( \lambda-a)^2\right] - 2 a \lambda,  \\
	\mathcal{D} &= \frac{4 \left(-1 + a^2 B^2\right) \left[\eta + ( \lambda-a)^2\right]}{-2 + a^2 B^2},\\
	\mathcal{G} &=-a^2\eta.
\end{aligned}
\eea

Due to the complex form of the radial potential,  a full analytical treatment is computationally intractable. Fortunately, 
as described in Eq.~\eqref{Bgauss}, in the scenario we consider, even a small value of the parameter $B$ can correspond to an extremely large physical magnetic field. It is reasonable to consider a small value of $B$ in the following study. Under this limit, we only retain the first-order corrections of the four roots of the radial potential with respect to the magnetic field. The roots of the radial potential in KBR spacetime can therefore be expressed based on the Kerr case as
\bea\label{eq:root4expan}
	r_j=r_j^k+B^2 \delta r_j,\quad j=1,2,3,4.
\eea
The roots of radial potential in the Kerr spacetime are given by \cite{Gralla:2019drh}, 
\bea
\begin{aligned}
		r_1^k = -z - \sqrt{-\frac{\mathcal{A^{'}}}{2} - z^2 + \frac{\mathcal{B^{'}}}{4z}},& \quad
		r_2^k = -z + \sqrt{-\frac{\mathcal{A^{'}}}{2} - z^2 + \frac{\mathcal{B^{'}}}{4z}}, \\
		r_3^k = z - \sqrt{-\frac{\mathcal{A^{'}}}{2} - z^2 - \frac{\mathcal{B^{'}}}{4z}},& \quad
		r_4^k = z + \sqrt{-\frac{\mathcal{A^{'}}}{2} - z^2 - \frac{\mathcal{B^{'}}}{4z}},
\end{aligned}
\eea
where
\bea
\begin{gathered}
z = \sqrt{\frac{\omega_+ + \omega_-}{2} - \frac{\mathcal{A^{'}}}{6}},\quad
\omega_{\pm} = \sqrt[3]{-\frac{\mathcal{Q}}{2} \pm \sqrt{\left( \frac{\mathcal{P}}{3} \right)^3 + \left( \frac{\mathcal{Q}}{2} \right)^2}},\\
\mathcal{P} = -\frac{\mathcal{A^{'}}^2}{12} - \mathcal{C^{'}}, \quad
\mathcal{Q} = -\frac{\mathcal{A^{'}}}{3}\left[\left( \frac{\mathcal{A^{'}}}{6} \right)^2 - \mathcal{C^{'}}\right] - \frac{\mathcal{B^{'}}^2}{8},\\
\mathcal{A^{'}} = a^2 - \eta^k - (\lambda^k)^2, \quad
\mathcal{B^{'}} = 2\left[\eta^k + (\lambda^k - a)^2\right], \quad
\mathcal{C^{'}} = -a^2\eta^k,  
\end{gathered}
\eea
with
\begin{equation}
	\lambda^k = a + \frac{r}{a} \left[ r - \frac{2(r^2-2r+a^2)}{r - 1} \right], \quad
	\eta^k = \frac{r^3}{a^2} \left[ \frac{4(r^2-2r+a^2)}{(r - 1)^2} - r \right].
\end{equation}
The correction term is given by
\begin{equation}
	\begin{aligned}
	\delta r_j &=\frac{r_j^k \left[a^2 (1 +r_j^k) + r_j^k \left(-1 - 2 r_j^k +(r_j^k)^2\right)\right] \left[\eta + ( \lambda-a)^2\right]}{2 \left[2 (r_j^k)^3 + a^2 (1 +r_j^k) - 2 a \lambda +  \left(\eta + \lambda^2\right) - r_j^k \left(\eta + \lambda^2\right)\right]}.
	 \end{aligned}
\end{equation}
In the KBR spacetime and for a small value of  $B$, we denote the four roots of the radial potential $R(r)$ by $r_1\leqslant r_2\leqslant r_3\leqslant r_4$ when they are real. In the near-critical regime, the two outer roots lie close to the photon-shell radius $\tilde{r}$. Outside the critical curve they are distinct and satisfy $r_+<r_3<r_4$. On the critical curve they merge into a double root $r_3=r_4$, while inside the curve they become a complex conjugate pair.
Consequently, the outermost real root $r_4$ (when $r_4>r_+$) is the unique radial turning point for any ray that escapes to infinity. These results reduce to the well-known Kerr case when $B=0$.

For the rays without a turning point, the radial integral $I_r$, $I_{\phi}$, and $I_t$ are a single-valued functions. However, for rays with a turning point, these radial integrals must be double-valued functions. Let $w=0, 1$ denote the direct rays and the reflected rays. Then the radial integrals can be written as 
\begin{equation}
	I_r=2w\int_{r_4}^{r_s}\frac{dr}{ \sqrt{\mathcal{R}(r)}}+\int_{r_s}^{r_o}\frac{dr}{ \sqrt{\mathcal{R}(r)}},
\end{equation}
where $r_4$ is the largest real root of the radial potential $\mathcal{R}(r)$.

\section{Critical rays and key parameters}\label{sec:paras}

Critical rays correspond to the orbits of photons that are extremely bent by a black hole, we impose the double root condition of the radial potential
\begin{equation}
	\mathcal{R}(\tilde{r}) = \partial_r \mathcal{R}(\tilde{r}) = 0.
\end{equation}
For $\tilde{r} > r_+$, solving these equations yields the critical impact parameters
\begin{align}
	\tilde{\lambda}(\tilde{r}) &= \frac{\tilde{r}^2 + a^2}{a} - \frac{4 \tilde{r} Q(\tilde{r})}{a Q'(\tilde{r})},\label{lambdar}\\
	\tilde{\eta}(\tilde{r}) &= \frac{16 \tilde{r}^2 Q(\tilde{r})}{Q'(\tilde{r})^2} - \left( \frac{\tilde{r}^2}{a} - \frac{4\tilde{r} Q(\tilde{r})}{a Q'(\tilde{r})} \right)^2. \label{eta}
\end{align}
Quantities bearing a tilde are corresponding to the relevant quantities associated with critical photon. From $\eta\geqslant0$, it can be inferred that $\tilde{r}$ is in the range of $\tilde{r}_{\text{p}} \leqslant \tilde{r} \leqslant \tilde{r}_{\text{m}}$, with $\tilde{r}_{\text{p,m}}$ being the solutions of $\eta(\tilde{r})=0$, corresponding to the prograde orbit and retrograde orbit, respectively. Similar to Eq. \eqref{eq:root4expan}, we can also expand the boundary $\tilde{r}_{\text{p,m}}$ with respect to the magnetic field around the Kerr case, that is
\begin{equation}
	\tilde{r}_{{\text{p,m}}} = \tilde{r}^k_\text{p,m} + B^2\delta \tilde{r}_{\text{p,m}},\label{p}
\end{equation}  
where  
\begin{equation}
	\tilde{r}^k_{\text{p,m}} = 2\left(1 + \cos\varphi_\pm\right), \quad \varphi_\pm = \frac{2}{3} \cos^{-1}(\pm a),
\end{equation}  
are the corresponding boundaries in the Kerr spacetime. Besides,
\begin{equation}
	\delta \tilde{r}_{\text{p,m}} =\frac{ (\cos^2\varphi_\pm + 1) \left[ 3a^4 + 48a^2 + 7(5a^2 + 4)\cos\varphi_\pm + (11a^2 + 26)\cos 2\varphi_\pm + 2\cos 4\varphi_\pm \right]}{3\left[a^2 + \cos\varphi_\pm + \cos 2\varphi_\pm\right]},
\end{equation}  
represents the first-order correction to the photon region due to the magnetic field in the KBR spacetime. For the polar observer, among all spherical photon orbits, only the orbit with a specific orbital radius $\tilde{r}_0$ can reach the observer, satisfying $\tilde{\lambda}(\tilde{r}_0)=0$. Its radius can also be obtained through the expansion
\begin{equation}
\tilde{r}_0=\tilde{r}_0^k+B^2\delta \tilde{r}_0\,,
\end{equation}
where $\tilde{r}_0^k$ is the orbit in the Kerr spacetime that can reach the polar axis
\begin{equation}
	\tilde{r}_0^k = 1 + 2\sqrt{1 - \frac{a^2}{3}} \cos\left[ \frac{1}{3} \arccos \frac{\left(1 - a\right)}{\left(1 - a^2\right)^{3/2}} \right],
\end{equation}
and
\begin{equation}
	\delta \tilde{r}_0=-\frac{\tilde{r}_0^k \left[a^4 (1 + 2\tilde{r}_0^k) + a^2 \tilde{r}_0^k \left(-2 - 9 \tilde{r}_0^k + 4 \tilde{r}{_0^k}^2 \right) + 2 \tilde{r}{_0^k}^2\left(1 + 4 \tilde{r}_0^k - 4 \tilde{r}{_0^k}^2  + \tilde{r}{_0^k}^3 \right)\right]}{2 a^2 - 2 \tilde{r}_0^k \left(3 - 3 \tilde{r}_0^k + \tilde{r}{_0^k}^2 \right)}.
\end{equation}

We have verified that for spherical photon orbits $\partial_r^2(\mathcal{R}(r))<0$, which implies that the orbits are unstable. The rate at which nearby orbits deviate can be characterized by the Lyapunov exponent $\gamma$, which is defined using the fractional number of orbits as a parameter \cite{Johnson:2019ljv}. Assume the photon travels from $r_1=\tilde{r}(1+\delta r_1)$ to $r_2=\tilde{r}(1+\delta r_2)$. Without loss of generality, let $r_1>r_2$ and the photon moves toward $\tilde{r}$. Since there is no turning point between $r_1$ and $r_2$, the radial integral can be written directly as
\begin{equation}
	I_r=\int_{r_1}^{r_2}\frac{dr}{\sqrt{\mathcal{R}(r)}}\,.
\label{I_r}
\end{equation}
By applying Eq.~\eqref{n} and Eq.~\eqref{I_r}, we can obtain that the radial integral exhibits a logarithmic divergence along the path between \(r_1\) and \(r_2\), namely,
\begin{equation}
	\frac{r_2-\tilde{r}}{r_1-\tilde{r}}\approx \exp[2 \gamma (n_2-n_1)],
\end{equation}
where $n_1$ and $n_2$ denote the fractional orbits at $r_1$ and $r_2$, respectively. The Lyapunov exponent $\gamma$ can be given by
\begin{equation}
	\gamma=\frac{4 \tilde{r}\sqrt{\mathcal{\tilde{X}}(\tilde{r})}}{\sqrt{\tilde{A_2}}\sqrt{-\tilde{u}_-}}\tilde{K},
\label{g1}	
\end{equation}
where $\tilde{K}=K(\tilde{u}_+/\tilde{u}_-)$ and $\mathcal{\tilde{X}}$ is defined as
 \begin{equation}
 	\mathcal{\tilde{X}}(\tilde{r})=1+\frac{2Q(\tilde{r})\left[Q'(\tilde{r})-\tilde{r} Q''(\tilde{r})\right]}{\tilde{r} Q'(\tilde{r})^2}.
 \end{equation}

Another key parameter $\delta$ is defined as the azimuthal angle change in half oscillation period in the $\theta$ direction. Substituting $\tilde{r}$ into Eq.~\eqref{I_{phi}} and Eq.~\eqref{lambdar}, we have 
\begin{equation}
	I_{\phi} \approx  \frac{4 a \tilde{r}}{Q'(\tilde{r})} I_r= \frac{4 a \tilde{r}}{Q'(\tilde{r})} G_{\theta}\label{I_{phi1}}.
\end{equation}
By applying Eq.~\eqref{I_{phi1}} and Eq.~\eqref{dphi},the variation of $\phi$ for a spherical orbit is expressed as
\begin{equation}
	\Delta \phi= \frac{4 a \tilde{r}}{Q'(\tilde{r})}G_{\theta}+G_{\phi}.
\end{equation}
Using the elliptic integral formulas Eq.~\eqref{a} and  Eq.~\eqref{b}, setting $\theta_{s}=\theta_{o}$ with $m=2$, we can obtain the azimuthal angle change within a half period,
\begin{equation}
	\hat{\delta}= \frac{2}{\sqrt{A_2}\sqrt{ -u_{-}}} \left[\frac{4 a \tilde{r}}{Q'(\tilde{r})}\tilde{K}+C_1 \tilde{\Pi}_1+(C_2-a) \tilde{\Pi}_2\right].
\label{delta1}
\end{equation}
In agreement with \cite{Teo:2003ltt}, the quantity $\hat{\delta} $ is not a smooth function of $r$ and develops a jump across the pole-crossing orbit $\tilde{r}=\tilde{r}_0$.  In the Kerr case, this jump equals $2\pi$ and can be captured by adding $2 \pi H(\tilde{r}-\tilde{r}_0)$.  
However, in the KBR spacetime the azimuthal coordinate is subject to an axis-regularity condition \cite{Wang:2025vsx}: requiring the absence of singularities at $\theta=0$ and $\pi$ yields $2\pi=\phi_{\text{max}}P(0)=\phi_{\text{max}}P(\pi)$, so the period of the azimuthal angle is $\phi_{\text{max}}=2 \pi/P(0)$. Since the pole-crossing discontinuity corresponds precisely to one azimuthal-period mismatch when passing the axis, the appropriate jump size for $\hat{\delta}$ in our convention is $\phi_{\text{max}}$ rather than $2 \pi$. Therefore we obtain the continuous version as
\begin{equation}
	\delta=\hat{\delta}+\frac{2 \pi}{1+C} H(\tilde{r}-\tilde{r}_0).
\label{delta2}
\end{equation}
Here $C$ is the parameter entering $P(\theta)=1+C\cos^2\theta$. Combining Eqs.~\eqref{delta1} and ~\eqref{delta2} gives
\begin{equation}
	\delta=\frac{2}{\sqrt{A_2}\sqrt{ -u_{-}}} \left[\frac{4 a \tilde{r}}{Q'(\tilde{r})}\tilde{K}+C_1 \tilde{\Pi}_1+(C_2-a) \tilde{\Pi}_2\right] + \frac{2 \pi}{1+C} H(\tilde{r}-\tilde{r}_0).\label{d1}
\end{equation}

The last key parameter $\tau$ is defined as the elapsed time over a half-oscillation period in the $\theta$ direction. Analogous to that used in Eq.~\eqref{I_{phi1}}, we find  
\begin{equation}
	I_t\approx\frac{4 \tilde{r} \left(a^2 + \tilde{r}^2\right)}{Q'(\tilde{r})}I_r=\frac{4 \tilde{r} \left(a^2 + \tilde{r}^2\right)}{Q'(\tilde{r})}G_{\theta}.
\label{1}
\end{equation}
Similarly, by applying Eq.~\eqref{deltat} and Eq.~\eqref{1}, the variation of $t$ for a spherical orbit can be expressed as
\begin{equation}
	\Delta t =\frac{4 \tilde{r} \left(a^2 + \tilde{r}^2\right)}{Q'(\tilde{r})}G_{\theta}+a G_t.
\end{equation}
Using the elliptic integral Eq.~\eqref{a} and Eq.~\eqref{Gt}, and setting $\theta_{s}=\theta_{o}$ with $m=2$, we can get the time-delay within a half-period as
\begin{equation}
	\tau=\frac{2}{\sqrt{A_2}\sqrt{ -u_{-}}}\left[\frac{4 \tilde{r} \left(a^2 + \tilde{r}^2\right)}{Q'(\tilde{r})}\tilde{K}+\frac{a^2}{C}\tilde{K}+\frac{Ca(\tilde{\lambda}-a)-a^2}{C}\tilde{\Pi}_2 \right].
\label{t1}
\end{equation}
We refer to $\gamma$, $\delta$ and $\tau$ as the critical parameters that characterize the critical behavior of unstable photon orbits. These parameters not only govern the behavior of critical photons but also determine the self-similar characteristics of gravitational lensing images. Analogous to Eq.~\eqref{eq:root4expan}, we can expand the three critical parameters around the Kerr case. For brevity, we present here only the expansion for $\gamma$ 
\begin{equation}
	\gamma  =\gamma^k+B^2\Delta \gamma,
\end{equation}
where $\gamma^k$ is the Lyapunov exponent in the Kerr spacetime, given by
\begin{equation}
	\gamma^k=\frac{4 \tilde{r} \sqrt{\mathcal{\tilde{X}}^k}}{a \sqrt{-u^k_-}}\tilde{K}^k,
\end{equation} 
where
\begin{equation}
	\tilde{K}^{k} = \tilde{K}^{k}\left( \frac{\tilde{u}^{k}_{+}}{\tilde{u}^{k}_{-}} \right) = \tilde{F}^{k}\left( \frac{\pi}{2} \bigg| \frac{\tilde{u}^{k}_{+}}{\tilde{u}^{k}_{-}} \right),
\end{equation}
with
\begin{equation}
	\mathcal{\tilde{X}}^k=1-\frac{\tilde{r}^2-2\tilde{r}+a^2}{\tilde{r}(\tilde{r}-1)^2},\quad
	\tilde{u}^k_{\pm} = \Delta_\theta \pm \sqrt{\Delta_\theta^2 + \frac{\tilde{\eta}^k}{a^2}}, \quad
	\Delta_\theta = \frac{1}{2}\left(1 - \frac{\tilde{\eta}^k + {\tilde{\lambda}^{k^2}}}{a^2}\right).
\end{equation}
The first-order corrections are given by
\begin{equation}
	\begin{aligned}
		\Delta\gamma=4 \tilde{r}&\left(\frac{\mathcal{\tilde{X}}}{2 a \sqrt{-\tilde{u}^{k}_{-}\mathcal{\tilde{X}}^{k}}}+\frac{-a^2\tilde{u}_-+(a^2-1)\tilde{u}^{k}_{-}(\tilde{\eta}^{k}+(a-\tilde{\lambda}^{k})^2) \sqrt{\mathcal{\tilde{X}}^{k}}}{2 a^3 \tilde{u}^{k}_{-}\sqrt{-\tilde{u}^{k}_{-}}} \right) \tilde{K}^{k} \\
		&+\frac{\pi (\tilde{u}^{k} \tilde{u}_+-\tilde{u}_-\tilde{u}^{k}_{+})\sqrt{\mathcal{\tilde{X}}^{k}}F_3}{8 (\tilde{u}^{k}_{-})^2 a \sqrt{-\tilde{u}^{k}_{-}}},
	\end{aligned}
\end{equation}
where  $F_3$ is a hypergeometric function, which can be expressed as
\begin{equation}
	F_3 = {}_2F_1\left(\frac{3}{2}, \frac{3}{2}; 2; \frac{u^{k}_{+}}{u^{k}_{-}}	\right).
\end{equation}
We defer the remaining two expansions to Appendix~\ref{app:expansion} due to their length.

\section{Key parameters results}\label{sec:keyparares}
In this section, we present the results for the three key parameters that we have computed. According to Eqs.~\eqref{g1}, ~\eqref{delta2}, and ~\eqref{t1}, the values of these parameters depend on the radius of the unstable circular photon orbit. For an on-axis observer, points on the photon ring correspond to the same orbital radius, whereas this is not the case for an off-axis observer. Therefore, we present these two situations separately.

\subsection{Off-axis observer}
Since the KBR spacetime is not asymptotically flat at infinity, we consider the Zero-Angular-Momentum Observers (ZAMOs), whose tetrad vectors are
\begin{equation}
	\hat{e}_0 = \zeta\, \partial_t + \Gamma\, \partial_\phi, \quad
	\hat{e}_1 = \frac{\partial_r}{\sqrt{g_{r r}}}, \quad
	\hat{e}_2 = \frac{\partial_\theta}{\sqrt{g_{\theta \theta}}}, \quad
	\hat{e}_3 = \frac{\partial_\phi}{\sqrt{g_{\phi \phi}}},
\end{equation}
where
\begin{equation}
	\begin{split}
		\zeta = \frac{g_{\phi \phi}}{\sqrt{g_{\phi \phi} \left(g_{\phi t}^2 - g_{\phi \phi} g_{t t}\right)}},  \qquad 
		\Gamma = \frac{-g_{\phi t}}{\sqrt{g_{\phi \phi} \left(g_{\phi t}^2 - g_{\phi \phi} g_{t t}\right)}}.
	\end{split}
\end{equation}
These basis vectors are normalized and mutually orthogonal. Then the four-momentum of a photon can be expressed as $p^{(\mu)} = p_\nu e^{\nu}_{(\mu)}$. To construct black hole images on the observational screen, we introduce the observation angles $(\alpha,\beta)$ which are determined by the tetrad components of the photon’s four-momentum, defined by
\bea
\cos \alpha = \frac{p^{(1)}}{p^{(0)}},\quad
\tan \beta = \frac{p^{(3)}}{p^{(2)}},
\eea
where
\begin{align}
	p^{(0)} &=p^{(t)}=\zeta-\lambda \Gamma, \\
	p^{(1)} &=p^{(r)}=\sqrt{\frac{\Omega^2(r,\theta)}{Q(r) \Sigma(r,\theta)}}\sqrt{\left[ (a^2+r^2)-a \lambda\right]^2-\left[ \eta+(\lambda-a)^2\right] Q(r)},\\
	p^{(2)} &=p^{(\theta)}=\sqrt{\frac{\Omega^2(r,\theta)}{P \Sigma(r,\theta)}} \sqrt{[\eta+(\lambda-a)^2]P-(a \sin \theta-\lambda \csc \theta)^2},\\
	p^{(3)} &=p^{(\phi)}=\frac{\lambda}{\sqrt{g_{\phi\phi}}}.
\end{align}
To obtain the relevant images of a black hole, it is further necessary to establish a correspondence between the observation angles $(\alpha,\beta)$ and the points $(x,y)$ on the imaging plane. We establish a standard Cartesian coordinates system, where the coordinates of points on this plane are given by the following equation
\begin{equation}
	x =-2r_o\tan (\frac{\alpha}{2})\sin \beta , \quad y =-2r_o\tan (\frac{\alpha}{2})\cos \beta, \label{coor}
\end{equation}
where $r_o$ is the distance from the observer to the black hole and we set $r_o=100$ in this work. Substituting the conserved quantities ~\eqref{lambdar} and ~\eqref{eta} into the Cartesian coordinates ~\eqref{coor}, and adopting the photon-shell radius range given by Eq.~\eqref{p}, we can obtain the critical curve.

\begin{figure}[htb!]
\centering
\includegraphics[width=\textwidth]{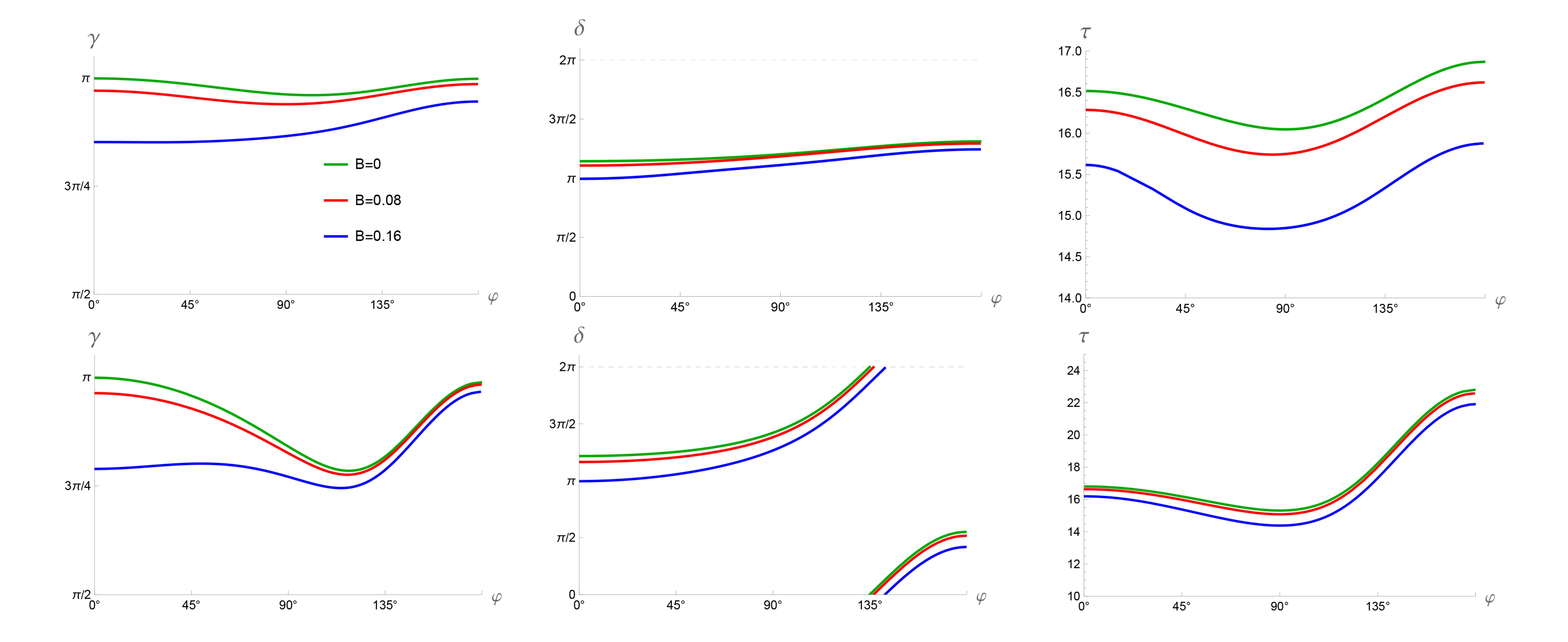}
\caption{Critical parameters $\gamma$, $\delta$ and $\tau$ evaluated along the critical curve of $\theta_o=80^{\circ}$, parameterized by the screen polar angle $\varphi$. Top row: $a=0.5$. Bottom row: $a=0.9$.}
\label{fig:offaxis80d}
\end{figure}

To describe the variation of the three key parameters along the critical curve, we first introduce the screen polar coordinates
\begin{equation}
		\rho=\sqrt{x^2+y^2}=2r_o\tan(\frac{\alpha}{2}),\quad \tan\varphi= \frac{y}{x}.
\end{equation}
Since the critical curve is symmetric with respect to the upper and lower halves, we can only consider the upper part, namely the region with $y>0$, corresponding to $0<\varphi<180^\circ$. Substituting the orbital radii $\tilde{r}$ corresponding to different $\varphi$ into Eqs.~\eqref{g1}, ~\eqref{delta2}, and ~\eqref{t1}, we obtain Fig.~\ref{fig:offaxis80d}, which illustrates the variation of the three parameters along the critical curve with inclination $\theta_o=80^{\circ}$ under different magnetic fields and black hole spins.

Overall, the case with smaller spin $a=0.5$ corresponds to a smoother variation of the parameters. In contrast, for the larger spin case $a=0.9$, the parameters vary more significantly along the critical curve. This behavior is related to the shape of the critical curve. When the observation inclination approaches $90^\circ$, a higher spin makes the critical curve more D-shaped, whereas a lower spin makes it closer to a circle. The variation of the three key parameters along a D-shaped curve is larger than along a nearly circular one.

\begin{figure}[htb!]
\centering
\includegraphics[width=\textwidth]{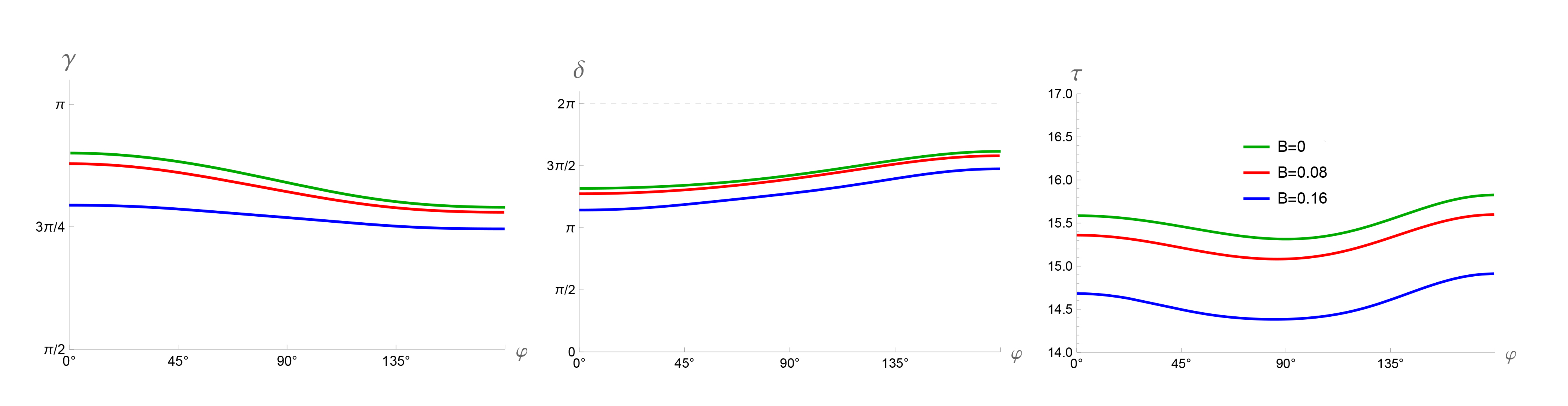}
\caption{Critical parameters evaluated along the critical curve of $a=0.9$ and $\theta_o=20^{\circ}$, parameterized by the screen polar angle $\varphi$.}
\label{fig:offaxis20d}
\end{figure}

This point can also be seen from the case with high spin but low inclination. In that situation, the critical curve is again close to circular. In Fig.~\ref{fig:offaxis20d} we show the results for a relatively large spin $a=0.9$ but a small inclination $\theta_o=20^\circ$, where the parameter variations are much smoother, which further supports the argument above.

Furthermore, in all of these figures, it can be seen that the magnetic field reduces the values of all three key parameters, which represents a characteristic difference between KBR black holes and Kerr black holes.

\subsection{On-axis observer}
For an on-axis observer, $g_{\phi\phi}=0$, which causes $\hat{e}_3$ to diverge. As a result, the previously defined Cartesian screen coordinates $(x,y)$ are no longer well defined. However, it is worth noting that $\rho = 2\tan(\alpha/2)$ does not depend on $\hat{e}_3$. This means that the radial screen coordinate $\rho$ can still be used. As mentioned above, for an on-axis observer the critical curve is a perfect circle, and all points on it correspond to the same photon orbital radius $\tilde{r}_0$. Therefore, the radius of the critical curve is given by
\begin{equation}
	\tilde{\rho}=2 r_o\sqrt{\frac{\Sigma(r_o,\theta_o)^2-\sqrt{\Sigma(r_o,\theta_o)^2-\left[ \tilde{\eta}(\tilde{r}_0)+a^2\right]Q(r_o) }}{\Sigma(r_o,\theta_o)^2+\sqrt{\Sigma(r_o,\theta_o)^2-\left[ \tilde{\eta}(\tilde{r}_0)+a^2\right]Q(r_o) }}}
\end{equation}
For face-on case, Eqs.~\eqref{g1}, ~\eqref{delta2}, and ~\eqref{t1} can be further simplified. Since $\lambda \to0$, we have
\begin{equation}
	u_+\to1, \quad u_-\to-\frac{\eta}{A_2} \label{limit1},
\end{equation}
and
\begin{equation}
	\theta_+\to\pi, \quad \theta_-\to0. \label{limit2}
\end{equation}
In particular, for the geodesic with $\eta>0$, each time the photon passes through the polar axis ($\theta=0$ or $\theta=\pi$), the angle $\phi$ undergoes a discontinuous change from $\phi\to\phi+\frac{\pi}{1+C}$. Mathematically, this is reflected in the divergence of the $\Pi_1$ term in the angular integral $G_{\phi}$ at the turning points (pole crossings). Thus, the second term of Eq.~\eqref{d1} can be written as 
\begin{equation}
	\lim_{\lambda \to 0^\pm} \frac{2 C_1 \Pi_1}{\sqrt{A_2} \sqrt{-u_-}}=\lim_{\lambda \to 0^\pm}\frac{1}{1+C}\frac{2\lambda \Pi_1}{\sqrt{A_2} \sqrt{-u_-}}=\pm \frac{ \pi}{1+C} . \label{eq:limphi}
\end{equation}
The critical parameters $\gamma$, $\delta$, and $\tau$ can now be given by
\begin{align}
	\gamma_0 &=\frac{4 \tilde{r}_0\sqrt{\mathcal{\tilde{X}}(\tilde{r}_0)}}{\sqrt{\tilde{\eta}(\tilde{r}_0)}}\tilde{K}\left(-\frac{\tilde{A_2}}{\tilde{\eta}(\tilde{r}_0)} \right), \\
	\delta_0 &=\frac{ \pi}{1+C} +\frac{2 a}{\sqrt{\tilde{\eta}(\tilde{r}_0)}}\left[ \frac{4\tilde{r}_0}{Q'(\tilde{r}_0)}\tilde{K}\left(-\frac{\tilde{A_2}}{\tilde{\eta}(\tilde{r}_0)} \right) -\Pi\left(-C\bigg\vert -\frac{\tilde{A_2}}{\tilde{\eta}(\tilde{r}_0)}\right) \right],  \\ 
	\tau_0  &=\frac{2}{\sqrt{\tilde{\eta}(\tilde{r}_0)}}\left[\frac{4\tilde{r}_0(\tilde{r}_0^2+a^2)}{Q'(\tilde{r}_0)}\tilde{K}\left(-\frac{\tilde{A_2}}{\tilde{\eta}(\tilde{r}_0)} \right)+\frac{a^2}{C}\tilde{K}\left(-\frac{\tilde{A_2}}{\tilde{\eta}(\tilde{r}_0)} \right)-\frac{a^2(1+C)}{C}\Pi\left(-C\bigg\vert -\frac{\tilde{A_2}}{\tilde{\eta}(\tilde{r}_0)}\right) \right].  
\end{align}
In the limits $B\to0$ and $a\to0$, the KBR spacetime reduces to the Schwarzschild spacetime, where the critical parameters are greatly simplified, giving $\gamma_0=\pi$, $\delta_0=\pi$, $\tau_0=3\sqrt{3}\pi$.

\begin{figure}[h]
	\centering
	\includegraphics[width=\textwidth]{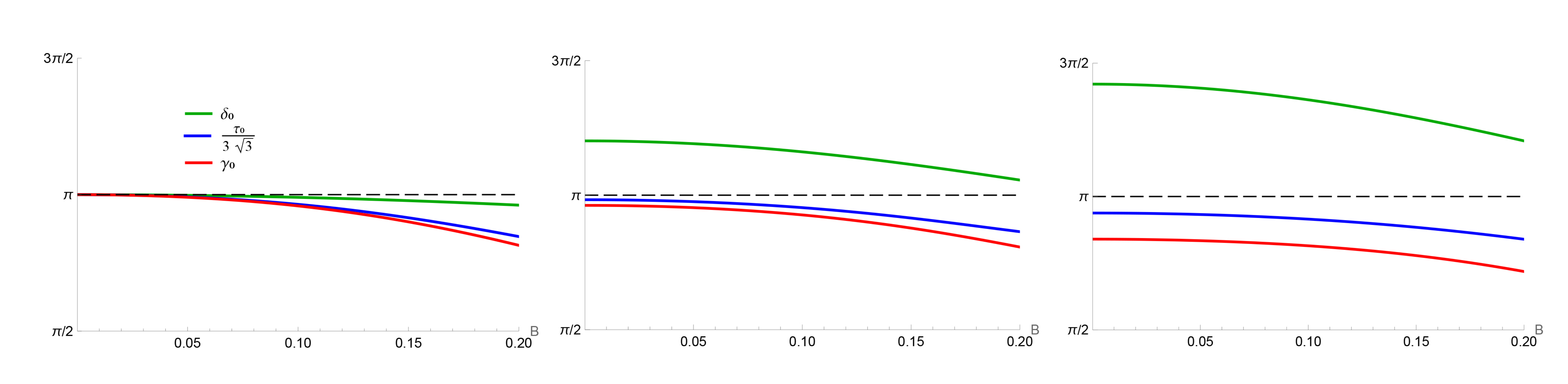}
	\caption{For an on-axis observer, the three parameters vary with the magnetic field strength for fixed black hole spin. From left to right, the panels correspond to $a=0$, $a=0.5$, and $a=0.9$, respectively. The green, blue, and red curves correspond to $\delta_0$, $\tau_0/3\sqrt{3}$, and $\gamma_0$, respectively.}
	\label{fig:fixspin}
\end{figure}

Fig.~\ref{fig:fixspin} shows the effect of the magnetic field on the three key parameters for three different values of the spin. From the figure, we can clearly see that all parameters decrease monotonically as the magnetic field increases, which is also consistent with the results for off-axis observers.

\begin{figure}[h]
	\centering
	\includegraphics[width=\textwidth]{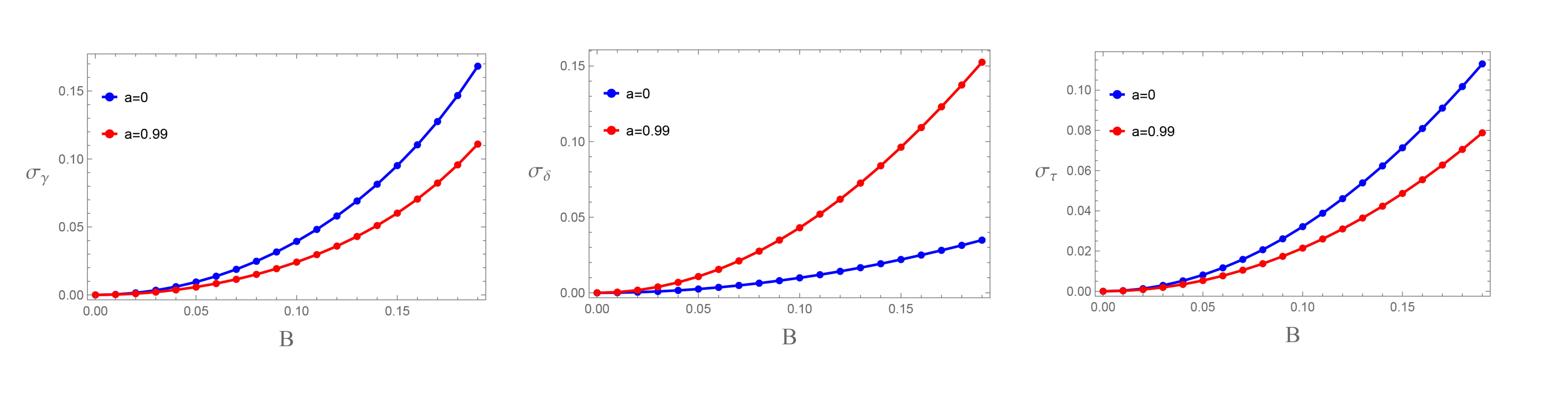}
	\caption{The deviations of the three parameters relative to the Kerr case. The blue and red curves correspond to the non-rotating case $a=0$ and the high-spin case $a=0.99$, respectively.}
	\label{devia}
\end{figure}

To quantitatively characterize the differences in critical parameters between the Kerr spacetime and the KBR spacetime, we introduce the dimensionless deviation $\sigma_\gamma$, $\sigma_\delta$ and $\sigma_\tau$, defined as
\begin{equation}
	\sigma_\gamma=1-\frac{\gamma_0}{\gamma_0(B=0)},\quad \sigma_\delta=1-\frac{\delta_0}{\delta_0(B=0)},\quad \sigma_\tau=1-\frac{\tau_0}{\tau_0(B=0)}.
\end{equation}
As shown in Fig.~\ref{devia}, the deviations increase monotonically with the magnetic-field strength $B$. To illustrate the effect of spin on these deviations, we plot two curves in each panel, corresponding to the non-rotating case $a=0$ and the near-extremal case $a=0.99$. The deviations for intermediate spin values lie between these two curves. We find that for $\sigma_\gamma$ and $\sigma_\tau$, higher spin leads to smaller deviations, whereas for $\sigma_\delta$ the opposite trend holds, with larger spin producing larger deviations, indicating a stronger spin dependence in the azimuthal angle change parameter. 

\begin{figure}[h]
	\centering
	\includegraphics[width=\textwidth]{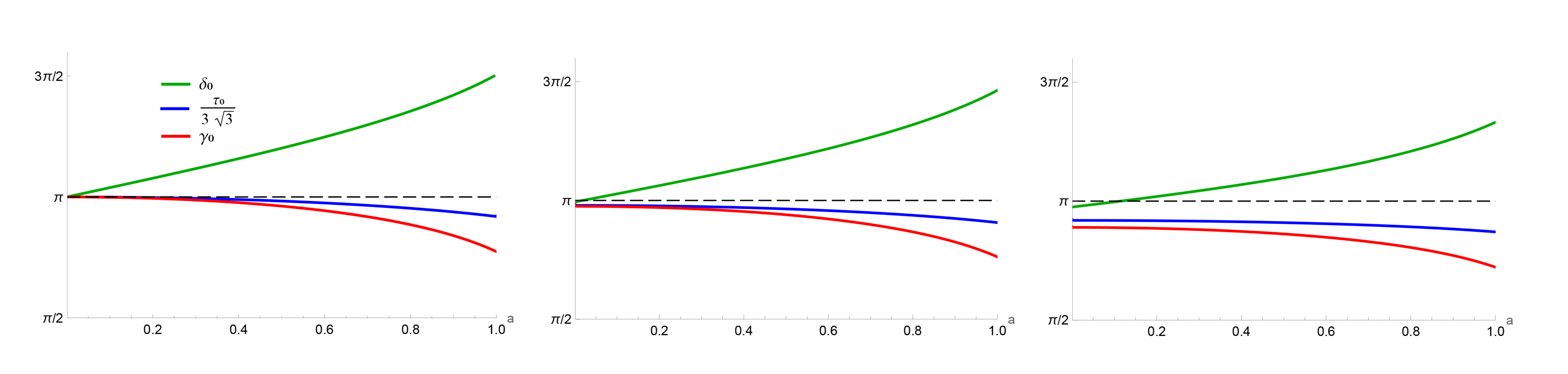}
	\caption{For an on-axis observer, the three parameters vary with the black hole spin for fixed magnetic field strength. From left to right, the panels correspond to $B=0$, $B=0.08$, and $B=0.16$, respectively. The green, blue, and red curves correspond to $\delta_0$, $\tau_0/3\sqrt{3}$, and $\gamma_0$, respectively.}
	\label{fig:fixB}
\end{figure}

In addition, Fig.~\ref{fig:fixB} shows the dependence of the three parameters on the spin for several fixed values of the magnetic field. In the presence of a magnetic field, the values of all three curves at $a=0$ are shifted downward to below $\pi$, satisfying $\gamma_0 < \tau_0/3\sqrt{3} < \delta_0 < \pi$. As the spin increases, $\delta_0$ grows monotonically and soon exceeds $\pi$, whereas $\gamma_0$ and $\tau_0/3\sqrt{3}$ both decrease monotonically and remain below $\pi$ throughout.

\section{Near-critical equations}\label{sec:lens}
In this section, we consider light rays near the photon ring and explore the recursive relations among higher-order images, thereby clarifying the physical meanings of the three key parameters.
First we introduce a nonnegative parameter $q=\sqrt{\eta}$. The impact parameters can be written as 
\begin{equation}
	\lambda=\tilde{\lambda}(1+\delta\lambda)  ,\quad q=\tilde{q}(1+\delta q),
\end{equation}
with $|\delta \lambda| \sim |\delta q|\ll 1$. We substitute it into the radial potential equation of the KBR spacetime and expand it near the critical radius $\tilde{r}$, thereby relating $\delta \lambda$ and $\delta q$ to the dimensionless quantity $\delta r^2_0$ (see Appendix \ref{app:radial})
 
\begin{equation}
	\delta r_0^2=\frac{ Q(\tilde{r})}{2 \tilde{r}^{\,4} \mathcal{\tilde{X}}(\tilde{r})} \left[   \tilde{q}^{\,2} \delta q + \tilde{\lambda} \left[  \frac{4 a \tilde{r}}{Q'(\tilde{r})} +   \left( \tilde{\lambda} - a \right) \right] \delta \lambda\right].
\end{equation}
Since $\delta r_0^2$ is positive/negative when located outside/inside the critical curve, we normalize it along the normal direction of the curve and define it as the signed normal distance $d$ to the critical curve on the screen
\begin{equation}
	d=\frac{\delta r_0^2}{K(\tilde{r})}=\frac{\delta r_0^2}{\sqrt{\left(\partial_x \delta r_0^2 \right)^2+\left(\partial_y \delta r_0^2 \right)^2}}.
\end{equation}
For $\delta r_0^2=0$, the case corresponds exactly to the critical curve, and the radial integral is given by Eqs.~\eqref{I_r},~\eqref{I_{phi1}} and ~\eqref{1}. For $\delta r_0^2 \neq 0$, the radial integral becomes
\begin{equation}
	I_r\approx-\frac{1}{2\tilde{r}\sqrt{\tilde{X}(\tilde{r})}}\log\left[ C_\pm d\right] \label{i},
\end{equation}
\begin{equation}
	I_{\phi}\approx\frac{4 a \tilde{r}}{Q'(\tilde{r})}I_r+D_{\pm} \label{j},
\end{equation}
\begin{equation}
	I_t\approx\frac{4 \tilde{r}(a^2+\tilde{r}^2)}{Q'(\tilde{r})}I_r+H_{\pm} \label{k},
\end{equation}
where $C_{\pm},D_{\pm}$ and $H_{\pm}$ are independent of $d$, and their expressions are given in Appendix~\ref{app:match}.

By using the Eqs.~\eqref{n} and~\eqref{n1}, together with Eqs.~\eqref{i}--\eqref{k} combined with the elliptical angular path integral~\eqref{a}--\eqref{Gt}, the lens equations ~\eqref{Mino}--\eqref{deltat} can be expressed as
\begin{align}
	d\approx\frac{1}{C_\pm}e^{-2n\gamma} \label{eq:d},\\
	\Delta \phi \approx 2n\hat{\delta} -J_{\pm}^{\phi} \label{eq:dp},\\
	\Delta t\approx 2 n \tau-J_{\pm}^t \label{eq:dt},
\end{align}
where 
\begin{equation}
	\begin{aligned}
	J_{\pm}^{\phi} &=\pm_{o}\left[ \frac{C_1 \tilde{\Pi}_1}{\sqrt{A_2}\sqrt{ -u_{-}}}\left[(-1)^m\left( \frac{\tilde{F}_s}{\tilde{K}}-\frac{\tilde{\Pi}_{1s}}{\tilde{\Pi}_1}\right) - \left( \frac{\tilde{F}_o}{\tilde{K}}-\frac{\tilde{\Pi}_{1o}}{\tilde{\Pi}_1}\right)\right]  \right] \\
	&\pm_{o}\left[ \frac{(C_2-a) \tilde{\Pi}_2}{\sqrt{A_2}\sqrt{ -u_{-}}}\left[(-1)^m\left( \frac{\tilde{F}_s}{\tilde{K}}-\frac{\tilde{\Pi}_{2s}}{\tilde{\Pi}_2}\right) - \left( \frac{\tilde{F}_o}{\tilde{K}}-\frac{\tilde{\Pi}_{2o}}{\tilde{\Pi}_2}\right)\right]  \right]-D_{\pm},
	\end{aligned}
\end{equation}
\begin{equation}
	J_{\pm}^t=\pm_{o}\frac{\left[ C a(\lambda -a)-a^2\right]\tilde{\Pi}_2 }{C \sqrt{A_2}\sqrt{ -u_{-}}}\left[(-1)^m\left( \frac{\tilde{F}_s}{\tilde{K}}-\frac{\tilde{\Pi}_{2s}}{\tilde{\Pi}_2}\right) - \left( \frac{\tilde{F}_o}{\tilde{K}}-\frac{\tilde{\Pi}_{2o}}{\tilde{\Pi}_2}\right) \right] -H_{\pm} .
\end{equation}
These near-critical equations determine the higher-order images. Next, based on the above-mentioned near-critical lens equations, we systematically discuss the characteristics of photon rings for different source positions, and consider two cases separately, polar-axis observers and off-axis observers. 
\subsection{Polar-axis Observer}

We first consider the simplest case: the equatorial source observed on the polar axis, for which we have $\theta_s=\pi/2$, $\tilde{F}_s=0$, $\tilde{\Pi}_{1s}=0$ and $\tilde{\Pi}_{2s}=0$. Eq.\eqref{n1} reduces to
\begin{equation}
	n\approx\frac{m}{2}+\frac{1}{4}.
\end{equation}
Accordingly, we label each near-critical (photon-ring) image with the integer $m$, and we then consider two successive images $m$ and $m+1$, from Eqs.\eqref{eq:d}--\eqref{eq:dt} we can get
\begin{align}
	\frac{d_{m+1}}{d_m} &\approx \mathrm{e}^{-\gamma_0},\\
	(\Delta \phi)_{m+1}-(\Delta \phi)_{m} &\approx \delta_0,\\
	(\Delta t)_{m+1}-(\Delta t)_{m} &\approx \tau_0.
\end{align}
Note that $C_{\pm}$, $D_{\pm}$ and $H_{\pm}$ are eliminated in this result.

For general source $(\theta_s\neq\pi/2)$,  Eq.\eqref{n1} reduces to
\begin{equation}
	n=\frac{m}{2}+\frac{1}{4}-\frac{(-1)^m}{4}\frac{\tilde{F}_s}{\tilde{K}}.
\end{equation}
We also consider two successive images $m$ and $m+1$
\begin{align}
	\frac{d_{m+1}}{d_m}&\approx \text{exp}\left[-\left[ 1+(-1)^m\frac{\tilde{F}_s}{\tilde{K}}\right] \gamma_0 \right], \\
	(\Delta \phi)_{m+1}-(\Delta \phi)_{m} &\approx \left[ 1+(-1)^m \frac{\tilde{F}_s}{\tilde{K}}\right]\delta_0-\frac{\pi}{1+C}\left[ (-1)^m\frac{\tilde{F}_s}{\tilde{K}}\right] + \frac{2a \tilde{\Pi}_2}{\sqrt{\tilde{\eta}(\tilde{r}_0)}}\left[(-1)^m\left( \frac{\tilde{F}_s}{\tilde{K}}-\frac{\tilde{\Pi}_{2s}}{\tilde{\Pi}_2}\right) \right] \label{eq:dphi} ,  \\
	(\Delta t)_{m+1}-(\Delta t)_{m} &\approx \left[ 1+(-1)^m \frac{\tilde{F}_s}{\tilde{K}}\right]\tau_0+\frac{ 2 a^2(1+C)\tilde{\Pi}_2 }{C \sqrt{\tilde{\eta}(\tilde{r}_0)}}\left[(-1)^m\left( \frac{\tilde{F}_s}{\tilde{K}}-\frac{\tilde{\Pi}_{2s}}{\tilde{\Pi}_2}\right) \right],
\end{align}
where we have use Eq.~\eqref{eq:limphi} in Eq.\eqref{eq:dphi} to handle the case of photons arriving at the axis.

\subsection{Off-axis Observer}
For an off-axis observer $\theta_o\neq0$, we have $\pm_{o}=\text{sign}(p^{\theta}_o)$ and the key parameters $\gamma$, $\delta$ and $\tau$ depend on $\tilde{r}$. For equatorial source $\theta_s=\pi/2$, the two successive images $m$ and $m+1$ also satisfy
\begin{align}
	\frac{d_{m+1}}{d_m} &\approx \mathrm{e}^{-\gamma},\\
	(\Delta \phi)_{m+1}-(\Delta \phi)_{m} &\approx \delta,\\
	(\Delta t)_{m+1}-(\Delta t)_{m} &\approx \tau.
\end{align}

For general source $(\theta_s\neq\pi/2)$, we have
\bea
\begin{aligned}
	\frac{d_{m+1}}{d_m}&\approx\text{exp}\left[-\left[1-\text{sign}(p^{\theta}_o) (-1)^m \frac{\tilde{F}_s}{\tilde{K}}\right]\gamma  \right] ,\\
	(\Delta \phi)_{m+1}-(\Delta \phi)_{m} &\approx \left[1-\text{sign}(p_o^{\theta})(-1)^m\frac{\tilde{F}_s}{\tilde{K}} \right] \delta +\text{sign}(p_o^{\theta})(-1)^m\frac{2C_1\tilde{\Pi}_1}{\sqrt{A_2}\sqrt{-u_-}}\left( \frac{\tilde{F}_s}{\tilde{K}}-\frac{\tilde{\Pi}_{1s}}{\tilde{\Pi}_1}\right) \\
	&\qquad +\text{sign}(p_o^{\theta})(-1)^m\frac{(C_2-a)\tilde{\Pi}_2}{\sqrt{A_2}\sqrt{-u_-}} \left( \frac{\tilde{F}_s}{\tilde{K}}-\frac{\tilde{\Pi}_{2s}}{\tilde{\Pi}_2}\right),\\
	(\Delta t)_{m+1}-(\Delta t)_{m} &\approx \left[1-\text{sign}(p_o^{\theta})(-1)^m\frac{\tilde{F}_s}{\tilde{K}} \right] \tau  \\
	 &\qquad+\text{sign}(p_o^{\theta})(-1)^m\frac{2\left[C a (\lambda-a)-a^2 \right]\tilde{\Pi}_2 }{C \sqrt{A_2}\sqrt{-u_-}}\left(\frac{\tilde{F}_s}{\tilde{K}} -\frac{\tilde{\Pi}_{2s}}{\tilde{\Pi}_2}\right) .
\end{aligned}
\eea
If we consider the $m$-th and $(m+2)$-th order images, the above equation can be greatly simplified
\bea
\begin{aligned}
\frac{d_{m+2}}{d_m}&\approx \mathrm{e}^{2\gamma}, \\
(\Delta \phi)_{m+2}-(\Delta \phi)_{m} &\approx 2 \delta, \\
(\Delta t)_{m+2}-(\Delta t)_{m} &\approx 2\tau.
\end{aligned}
\eea
These second-order recursive relations are applicable to arbitrary source configurations for any observer, including the special on-axis case mentioned earlier.

\section{Summary}\label{sec:sum}

In this paper, we employed a Kerr–Bertotti–Robinson (KBR) black hole with a background magnetic field as the theoretical model to investigate how the magnetic field altered the fine structure of photon rings and their higher-order images. We used the photon geodesic equations to analyze the behavior of radial and angular motion under the small magnetic field approximation. We then focused on unstable circular orbits and calculated three key parameters that characterize the near-ring critical behavior, $\gamma$, $\delta$, and $\tau$, which describe radial compression, azimuthal advancement, and time delay, respectively.

In the presence of a background magnetic field, the three critical parameters $\gamma$, $\delta$, and $\tau$ exhibit a systematic decrease compared to the non-magnetized Kerr case. This trend indicates that the originally pronounced self similar hierarchical structure of higher order photon rings is globally weakened, which enhances the potential observability of their fine structure. More specifically, a smaller $\gamma$ increases the radial separation between adjacent subrings, a smaller $\delta$ suppresses azimuthal phase advancement and reduces the relative rotation in position angle, and a smaller $\tau$ shortens the time delay scale between successive image orders.

With the anticipated improvement in observational precision from the next-generation Event Horizon Telescope, robust measurements of these fine-structure features can be directly compared with theoretical predictions. Such comparisons would enable tighter constraints on the black hole spin and related parameters, and also provide a way to infer the effective magnetic field strength $B$ in the near horizon region.

\section*{Acknowledgments}
We thank Zelin Zhang for insightful discussions. The work is partly supported by NSFC Grant No. 12275004, No.12547123, No. 12547127, and No. 12588101.

\appendix

\section{Expansion for critical parameters}\label{app:expansion}

We can expand the remaining two parameters into the first-order correction form with respect to the magnetic field, that is
\bea
\begin{aligned}
  \delta& = \delta^{k}+B^2\Delta \delta,\\
  \tau &= \tau^k+B^2\Delta \tau.
\end{aligned}
\eea
The expressions of the remaining two parameters in the Kerr case are
\bea
\begin{aligned}
 		\delta^k&=\frac{2}{\sqrt{-\tilde{u}^{k}_{-}}} \left( \left( \frac{\tilde{r} + 1}{\tilde{r} - 1} \right) \tilde{K}^{k} + \frac{\tilde{\lambda}\tilde{\Pi}^{k}}{a} \right),\\
\tau^k&= \frac{2}{a\sqrt{-\tilde{u}^{k}_{-}}} \left( \tilde{r}^2 \left( \frac{\tilde{r} + 3}{\tilde{r} - 1} \right) \tilde{K}^k - 2a^2 \tilde{u}^{k}_{+} \tilde{E}' \right),
\end{aligned}
\eea
where
\[
\begin{gathered}
  	\tilde{\Pi}^{k} = \tilde{\Pi}^{k}\left( \tilde{u}^{k}_{+} \bigg| \frac{\tilde{u}^{k}_{+}}{\tilde{u}^{k}_{-}} \right) = \tilde{\Pi}^{k}\left( \tilde{u}^{k}_{+}; \frac{\pi}{2} \bigg| \frac{\tilde{u}^{k}_{+}}{\tilde{u}^{k}_{-}} \right), \\
  	\tilde{E}' =\tilde{E}'\left( \frac{\tilde{u}^{k}_{+}}{\tilde{u}^{k}_{-}} \right) = \tilde{E}'\left( \frac{\pi}{2} \bigg| \frac{\tilde{u}^{k}_{+}}{\tilde{u}^{k}_{-}} \right).\\
\end{gathered}
\]
The first-order corrections are given by
 
  \[
  \begin{aligned}
  	\Delta \delta &= \frac{	\left(- a^2 \tilde{u}_- 	+ ( a^2-1)\tilde{u}^{k}_{-}\big(\tilde{\eta}^{k} + (a - \tilde{\lambda}^{k})^2\big)	\right)	\left(	- a\tilde{K}^{k} 	+ \frac{2 a \tilde{K}^{k} \tilde{r}}{\tilde{r}-1} 	+ \tilde{\lambda}^{k} \tilde{\Pi}^{k}	\right)}{ a^3 \tilde{u}^{k}_{-}\sqrt{- \tilde{u}^{k}_{-}}}
  	\\
  	&\qquad+ \frac{2}{\sqrt{-a^2 \tilde{u}^{k}_{-}}}\Bigg(
  	- a \Bigg(
  	\frac{1}{4}( a^2-1) F_1 \pi\, \tilde{u}^{k}_{+}
  	+ \frac{F_3 \pi (\tilde{u}^{k}_{-}\tilde{u}_- - \tilde{u}_- \tilde{u}^{k}_{+})}{8 (\tilde{u}^{k}_{-})^2}
  	\Bigg)
  	\\
  	&\qquad
  	+ 4 a \tilde{r} \Bigg(
  	\frac{\tilde{K}^{k}(-a^2 + 2\tilde{r} - 2a^2\tilde{r} + 6\tilde{r}^2 - 4\tilde{r}^3)}{4( \tilde{r}-1)^2}	+ \frac{F_3 \pi (\tilde{u}^{k}_{-}\tilde{u}_- - \tilde{u}_- \tilde{u}^{k}_{+})}{16( \tilde{r}-1)(\tilde{u}^{k}_{-})^2}\Bigg)
  	\\
  	&\qquad
  	+ (1 -a^2)\tilde{K}^{k} \tilde{\lambda}^{k}	+ (\tilde{\lambda} + ( a^2-1)\tilde{\lambda}^{k})\tilde{\Pi}^{k}+ \frac{	\tilde{\lambda}^{k}(\tilde{u}^{k}_{-}\tilde{u}_- - \tilde{u}_- \tilde{u}^{k}_{+})
  	}{	2 (\tilde{u}^{k}_{-})^2	\left(\tilde{u}^{k}_{+} - \frac{\tilde{u}^{k}_{+}}{\tilde{u}^{k}_{-}}	\right)	}\left(\frac{\tilde{E}^{k}}{ \frac{\tilde{u}^{k}_{+}}{\tilde{u}^{k}_{-}}-1}+ \tilde{\Pi}^{k}\right)
  	\\
  	&\qquad
  	+ \frac{\tilde{\lambda}^{k}\tilde{u}^{k}_{+}}{	2(\tilde{u}^{k}_{+}-1)	\left(-\tilde{u}^{k}_{+} + \frac{\tilde{u}^{k}_{+}}{\tilde{u}^{k}_{-}}	\right)}\left(\tilde{E}^{k}+ \frac{	\tilde{K}^{k}(-\tilde{u}^{k}_{+} - \frac{\tilde{u}^{k}_{+}}{\tilde{u}^{k}_{-}})}{	\tilde{u}^{k}_{+}}+ \frac{	\frac{\tilde{u}^{k}_{+}}{\tilde{u}^{k}_{-}} + (\tilde{u}^{k}_{-})^2	}{	\tilde{u}^{k}_{+}}\tilde{\Pi}^{k}\right)
  	\Bigg),
  \end{aligned}
  \]
 
 \[ 
 \begin{aligned}
 	\Delta\tau &=\frac{1}{a^3 \tilde{u}^{k}_{-} \sqrt{-\tilde{u}^{k}_{-}}}
 	\left(-a^2 \tilde{u}_-  + \left(a^2-1\right) \tilde{u}^{k}_{-} \left( \tilde{\eta}^k + \left(a - \tilde{\lambda}^k\right)^2 \right) \right)
 	\\
 	&\qquad \times \left( \frac{2 \tilde{K}^k \tilde{r} \left(a^2 + \tilde{r}^2\right)}{ \tilde{r}-1}
 	+ \frac{1}{4} a^2 F_1 \pi \tilde{u}^{k}_{+} + a \tilde{K}^k \left(  \tilde{\lambda}^k -a\right)
 	\right)
 	\\
 	& \qquad+\frac{8 \tilde{r} \left(a^2 + \tilde{r}^2\right) }{a \sqrt{-\tilde{u}^{k}_{-}}} \left(
 	\frac{\tilde{K}^k \left(-a^2 + 2\tilde{r} - 2a^2\tilde{r} + 6\tilde{r}^2 - 4\tilde{r}^3\right)	}{	4\left( \tilde{r}-1\right)^2	}	+	\frac{	F_3 \pi \left( \tilde{u}^{k}_{-} \tilde{u}_+ - \tilde{u}_- \tilde{u}^{k}_{+} \right)}{	16\left( \tilde{r}-1\right) \left( \tilde{u}^{k}_{-} \right)^2}\right)
 	\\
 	& \qquad +\frac{2}{\sqrt{-\tilde{u}^{k}_{-}}}
 	\left(\tilde{K}^k \tilde{\lambda}+\left(\frac{1}{4} \left( a^2-1\right) F_1 \pi \tilde{u}^{k}_{+} + \frac{F_3 \pi \left( \tilde{u}^{k}_{-} \tilde{u}_+ - \tilde{u}_- \tilde{u}^{k}_{+} \right)}{8 \left( \tilde{u}^{k}_{-} \right)^2}\right)\left(  \tilde{\lambda}^k -a\right)\right),
 \end{aligned}
 \]
 
where $F_1$ is hypergeometric function, which can be expressed as 
\[
F_1 = {}_2F_1\left(
  	\frac{1}{2}, \frac{3}{2}; 2; \frac{u^{k}_{+}}{u^{k}_{-}}
  	\right).
\]

\section{Radial integrals for near-critical rays}\label{app:radial}
In this appendix, we derive the path integrals for near-critical rays in the KBR spacetime. The analysis of photons asymptotically approaching a critical orbit is facilitated by dividing their trajectory into two distinct regions based on the radial coordinate $r$ relative to the critical radius $\tilde{r}$: one originating from the vicinity of the photon orbit, and the other from the distant region. This division allows us to employ the method of matched asymptotic expansions, whereby the complex global integral is decomposed into two more tractable parts. We utilize this framework to compute the asymptotic approximations for the radial integrals.

To facilitate the computation, we introduce a nonnegative parameter
\begin{equation}
q=\sqrt{\eta}>0,
\end{equation}
for near-critical rays. The conserved quantities $\lambda$ and $q$ can be written as
\begin{equation}
\lambda=\tilde{\lambda}(1+\delta\lambda), \quad q=\tilde{q}(1+\delta q),
\end{equation}
with $|\delta\lambda| \sim |\delta q| \ll 1$. We employ a radial coordinate $\delta r$ defined by
\begin{equation}
r=\tilde{r}(1+\delta r).
\end{equation}
This allows us to distinguish two regions:
\begin{equation}
\text{Near}: \quad |\delta\lambda| \sim |\delta q| \sim |\delta r| \ll 1
\end{equation}
\begin{equation}
\text{Far}: \quad |\delta\lambda| \sim |\delta q| \ll |\delta r|
\end{equation}
These two asymptotic regimes overlap in the region where $|\delta\lambda| \sim |\delta q| \ll |\delta r| \ll 1$. Note that the near zone and far zone we emphasize refer to the distance from the photon orbit radius $\tilde{r}$. Moreover, the far zone is disjoint, consisting of a “right zone” at a finite distance (the observer’s position $r_o$) and a “left zone” that includes the event horizon \cite{Gralla:2019drh}.

In the near region, the radial potential $\mathcal{R}(r)$ can be expanded as
\begin{equation}
\mathcal{R}(r) = \frac{1}{2} \mathcal{R}''(\tilde{r}) \delta r^2 + \mathcal{R}_\lambda \delta \lambda + \mathcal{R}q \delta q,
\end{equation}
where
\begin{align}
\mathcal{R}''(\tilde{r}) &=\tilde{r}^2\left[ 8\tilde{r}^2+4(\tilde{r}^2+a^2-a\tilde{\lambda})-Q''(\tilde{r})[\tilde{q}^2+(\tilde{\lambda}-a)^2]\right], \\
\mathcal{R}\lambda &= \tilde{\lambda} \left[ - 8 a \tilde{r} \frac{Q(\tilde{r})}{Q'(\tilde{r})} - 2 Q(\tilde{r}) \left( \tilde{\lambda} - a \right) \right],\\
\mathcal{R}_q &=-2\tilde{q}^2 Q(\tilde{r}).
\end{align}
It is convenient to define
\begin{equation}
\mathcal{R}(r)=\mathcal{R}n(\delta r) = \frac{1}{2} \mathcal{R}''(\tilde{r}) \delta r^2 + \mathcal{R}\lambda \delta \lambda + \mathcal{R}_q \delta q = 4 \tilde{r}^{4}\mathcal{\tilde{X}}(\tilde{r}) \left( \delta r^2 - \delta r_0^2 \right), \label{near}
\end{equation}
where
\begin{align}
\delta r_0^2 & = \frac{ Q(\tilde{r})}{2 \tilde{r}^{,4} \mathcal{\tilde{X}}(\tilde{r})} \left[ \tilde{q}^{,2} \delta q + \tilde{\lambda} \left( \frac{4 a \tilde{r}}{Q'(\tilde{r})} + \tilde{\lambda} - a \right) \delta \lambda \right], \label{delta r_0} \\
\mathcal{\tilde{X}}(\tilde{r}) &=1+\frac{2Q(\tilde{r})\left[Q'(\tilde{r})-\tilde{r} Q''(\tilde{r})\right]}{\tilde{r} Q'(\tilde{r})^2}.
\end{align}

In the far region, the slight deviations of the photon's conserved quantities $(\lambda,\eta)$ from their critical values have a negligible influence on the radial potential $\mathcal{R}(r)$. The potential is therefore well approximated by its form at the critical point, obtained by fixing $\lambda=\tilde{\lambda}$ and $q=\tilde{q}$. The radial potential becomes
\begin{equation}
\mathcal{R}(r)=\mathcal{R}_f(\delta r) =4 \tilde{r}^{,4}\mathcal{\tilde{X}}(\tilde{r})\delta r^2 \mathcal{F}(\delta r), \label{far}
\end{equation}
where
\begin{equation}
\mathcal{F}(\delta r)=1+\delta r f_1 +\frac{\delta r^2 f_2}{4},
\end{equation}
\begin{equation}
f_1=\frac{1}{\mathcal{\tilde{X}}(\tilde{r})}\left( 1-\frac{2 \tilde{r} Q(\tilde{r})Q'''(\tilde{r})}{3 Q'(\tilde{r})^2}\right),
\end{equation}
\begin{equation}
f_2=\frac{1}{ \mathcal{\tilde{X}}(\tilde{r})}\left(1-\frac{2 \tilde{r}^2 Q(\tilde{r})Q''''(\tilde{r})}{3 Q'(\tilde{r})^2} \right).
\end{equation}

In the KBR spacetime, a physical light ray never encounters a radial turning point in the far region; therefore, the far-zone approximation~\ref{far} for the radial potential satisfies $\mathcal{R}_f(\delta r)>0$ along any null geodesic that reaches the distant observer. Radial turning points can only occur in the near region, where the potential is well approximated by the near-zone form $\mathcal{R}_n(\delta r)$. As discussed below Eq.~\ref{near}, when $\delta r_0^2>0$, the equation $\mathcal{R}_n( \delta r)=0$ has two real roots $\delta r=\pm \delta r_0$ close to the photon-shell radius $\tilde{r}$. The outer root $r_t=\tilde{r}(1+\delta r_0)$ represents the unique radial turning point of any ray that escapes back to infinity. When $\delta r_0^2<0$, the near-zone potential has no real zeros and the trajectory is monotonic in $r$, so the corresponding rays have no turning point. In the boundary case $\delta r_0^2=0$, the near-zone potential has a double root at $\delta r=0$, corresponding to critical null geodesics that asymptotically approach the photon shell and project onto the critical curve on the observer’s screen.

\section{Matched asymptotic expansion computations}\label{app:match}
In the near region, the radial integrals simplify to
\begin{align}
I_r^n(\delta r_a,\delta r_b) &=\frac{1}{2 \tilde{r}\sqrt{\tilde{\mathcal{X}}(\tilde{r})}}\int_{\delta r_a}^{\delta r_b}\frac{d(\delta r)}{\sqrt{\delta r^2-\delta r_0^2}},\\
I_{\phi}^n(\delta r_a,\delta r_b) & =\frac{4 a \tilde{r}}{Q'(\tilde{r})}I_r^n(\delta r_a,\delta r_b),\\
I_t^n(\delta r_a,\delta r_b)& = \frac{4 \tilde{r}(a^2+\tilde{r}^2)}{Q'(\tilde{r})}I_r^n(\delta r_a,\delta r_b),
\end{align}
where we use $n$ to denote the near region, and $\delta r_a$, $\delta r_b$ denote the $\delta r$ coordinate values corresponding to the radii $r_a$ and $r_b$, with $|\delta r_a|<|\delta r_b| \ll 1$.

For $\delta r>\delta r_0>0$, where photons arrive outside $\mathcal{C}$, the result for the radial integrals is $I_{r,\text{out}}^{n}(\delta r_a,\delta r_b)$. For $\delta r_0^2<0$, where photons arrive inside $\mathcal{C}$, the result for the radial integrals is $I_{r,\text{in}}^{n}(\delta r_a,\delta r_b)$. These results can be combined into a single formula:
\begin{equation}
I_r^{n}(\delta r_a,\delta r_b)=\frac{\text{sign}(\delta r)}{2 \tilde{r}\sqrt{\tilde{\mathcal{X}}(\tilde{r})}}\left.\log\left( \frac{|\delta r| + \sqrt{\delta r^2 - \delta r_0^2}}{|\delta r_0|} \right)\right|_{\delta r_a}^{\delta r_b}.
\end{equation}

In the far region, the radial integrals simplify to
\begin{align}
I_r^f(\delta r_a,\delta r_b) &=\frac{1}{2 \tilde{r}\sqrt{\tilde{\mathcal{X}}(\tilde{r})}}\int_{\delta r_a}^{\delta r_b}\frac{d(\delta r)}{\sqrt{\delta r^2 \mathcal{F}(\delta r)}},
\end{align}
where we use $f$ to denote the far region, with $|\delta\lambda| \sim |\delta q| \ll |\delta r_a|<|\delta r_b|$. We then obtain
\begin{equation}
I_r^f(\delta r_a,\delta r_b)=\frac{\text{sign}(\delta r)}{2\tilde r \sqrt{\mathcal{\tilde{X}}(\tilde{r})}}\left.\log\left[\frac{\sqrt{\mathcal{F}(\delta r)} + 1 + \frac{f_1}{2}\delta r}{\delta r}\right]\right|_{\delta r_a}^{\delta r_b},
\end{equation}
\begin{align}
I_{\phi}^f(\delta r_a,\delta r_b)& =\frac{4 a \tilde{r}}{Q'(\tilde{r})}I_r^f(\delta r_a,\delta r_b)+D(\delta r_a,\delta r_b),\\
I_{t}^f(\delta r_a,\delta r_b) &= \frac{4 \tilde{r}(a^2+\tilde{r}^2)}{Q'(\tilde{r})}I_r^f(\delta r_a,\delta r_b)+H(\delta r_a,\delta r_b),
\end{align}
where
\begin{equation}
D(\delta r_a,\delta r_b)=\frac{\text{sign}(\delta r)}{2 \tilde{r}\sqrt{\mathcal{\tilde{X}}(\tilde{r})}}\int_{\delta r_a}^{\delta r_b}\frac{G_{\phi}(r)-G_{\phi}(\tilde{r})}{|\delta r|\sqrt{\mathcal{F}(\delta r)}}d(\delta r),
\end{equation}
\begin{equation}
H(\delta r_a,\delta r_b)=\frac{\text{sign}(\delta r)}{2 \tilde{r}\sqrt{\mathcal{\tilde{X}}(\tilde{r})}}\int_{\delta r_a}^{\delta r_b}\frac{G_{t}(r)-G_{t}(\tilde{r})}{|\delta r|\sqrt{\mathcal{F}(\delta r)}}d(\delta r),
\end{equation}
and
\begin{equation}
G_{\phi}(r)=\frac{a(r^2+a^2-a\lambda)}{Q(r)},
\end{equation}
\begin{equation}
G_t(r)=\frac{(r^2+a^2)(r^2+a^2-a\lambda)}{Q(r)}.
\end{equation}

We define $H(\delta r_a,\delta r_b)$ and $D(\delta r_a,\delta r_b)$ as the finite remainders after subtracting the part proportional to $I_r^f(\delta r_a,\delta r_b)$. Due to the complexity of the expansion process and the fact that these terms are eliminated when calculating the lens equation, we do not present their complete expressions.

We use the matched asymptotic expansion method to calculate the radial integrals of photons. The total radial integral is obtained by connecting the near-region integral and the far-region integral through their overlapping region. We select an arbitrary matching radius $\delta R$ in the overlapping region ($|\delta\lambda| \sim |\delta q| \ll |\delta R| \ll 1$), and split the integral into one part from $\delta r_a$ to $\delta R$ and the remaining part from $\delta R$ to $\delta r_b$.

For photons reaching regions outside the critical curve, with $0<\delta r_0<\delta r_a \ll 1$ and $\delta r_a\ll\delta r_b$, we denote their radial integral by $I^{nf}_{r,\text{out}}(\delta r_a,\delta r_b)$:
\begin{equation}
\begin{aligned}
I^{nf}_{r,\text{out}}(\delta r_a,\delta r_b) &=I_r^n(\delta r_a,\delta R)+I^f_r(\delta R,\delta r_b)\\
&=\frac{1}{2 \tilde{r}\sqrt{\mathcal{\tilde{X}}(\tilde{r})}}\left[ \log\left. \left( \frac{|\delta r| + \sqrt{\delta r^2 - \delta r_0^2}}{|\delta r_0|} \right) \right|_{\delta r_a}^{\delta R} +\left.
\log\left(\frac{\sqrt{\mathcal{F}(\delta r)} + 1 + \frac{f_1}{2}\delta r}{\delta r}\right)\right|_{\delta R}^{\delta r_b} \right] \\
&=\frac{1}{2 \tilde{r}\sqrt{\mathcal{\tilde{X}}(\tilde{r})}}\left[-\frac{1}{2}\log\left(\frac{f_1^2-f_2 }{64}\delta r_0^2 \right) -\operatorname{arctanh} \left( \frac{\sqrt{\delta r_a^2 - \delta r_0^2}}{\delta r_a} \right) -\operatorname{arctanh}\left( \frac{\sqrt{\mathcal{F}(\delta r_b)}}{1+\frac{f_1 \delta r_b}{2}}\right) \right].
\end{aligned}
\end{equation}

In the final calculation, using the matched-asymptotic approximation, the arbitrary matching radius $\delta R$ disappears from the final expression. In the KBR spacetime, the far-region contributions along $\phi$ and $t$ can be written as
\begin{equation}
I_{\phi, \text{out}}^{nf}(\delta r_a,\delta r_b)=\frac{4 a \tilde{r}}{Q'(\tilde{r})}I_{r,\text{out}}^{nf}(\delta r_a,\delta r_b)+D(0,\delta r_b),
\end{equation}
\begin{equation}
I_{t,\text{out}}^{nf}(\delta r_a,\delta r_b)=\frac{4 \tilde{r}(a^2+\tilde{r}^2)}{Q'(\tilde{r})}I_{r,\text{out}}^{nf}(\delta r_a,\delta r_b)+H(0,\delta r_b),
\end{equation}
where $D(0,\delta r_b)$ and $H(0,\delta r_b)$ are the finite remainders obtained after factoring out the logarithmic term proportional to $I_r$. They are expressed with $\delta r=0$ as a unified reference point and are independent of the matching radius $\delta R$. More precisely, the matched result yields $D(\delta R,\delta r_b)$ and $H(\delta R,\delta r_b)$, which reduce to $D(0,\delta r_b)$ and $H(0,\delta r_b)$ up to $\mathcal{O}(\delta R)$ corrections. Importantly, this does not correspond to arbitrarily setting $\delta R\to 0$; the $\mathcal{O}(\delta R)$ contributions are higher-order under the matched-asymptotic expansion and are therefore omitted.

For photons that reach within the critical curve, we split the integral into two parts: the radial integral from the far region to the near region on the right side, with $0<\delta r_a\ll 1$ and $\delta r_a\ll\delta r_b$, denoted by $I_{r,\text{in}}^{nf}(\delta r_a,\delta r_b)$; and the integral from the far region to the near region on the left side, which includes the event horizon, with $\delta r_b<0$, $\delta r_a<0$, and $|\delta r_b|\ll|\delta r_a|$, denoted by $I_{r,\text{in}}^{fn}(\delta r_a,\delta r_b)$:
\begin{equation}
I_{r,\text{in}}^{nf}(\delta r_a,\delta r_b)=\frac{1}{2 \tilde{r}\sqrt{\mathcal{\tilde{X}}(\tilde{r})}}\left[-\frac{1}{2}\log\left(\frac{f_1^2-f_2 }{64}\delta r_0^2 \right) -\operatorname{arctanh} \left( \frac{\delta r_a}{\sqrt{\delta r_a^2 - \delta r_0^2}} \right) -\operatorname{arctanh}\left( \frac{\sqrt{\mathcal{F}(\delta r_b)}}{1+\frac{f_1 \delta r_b}{2}}\right) \right],
\end{equation}
\begin{equation}
I_{r,\text{in}}^{fn}(\delta r_a,\delta r_b)=\frac{1}{2 \tilde{r}\sqrt{\mathcal{\tilde{X}}(\tilde{r})}}\left[-\frac{1}{2}\log\left(\frac{f_1^2-f_2 }{64}\delta r_0^2 \right) +\operatorname{arctanh} \left( \frac{\delta r_b}{\sqrt{\delta r_b^2 - \delta r_0^2}} \right) -\operatorname{arctanh}\left( \frac{\sqrt{\mathcal{F}(\delta r_a)}}{1+\frac{f_1 \delta r_a}{2}}\right) \right].
\end{equation}

The integrals along the $\phi$ and $t$ directions can be expressed as follows:
\begin{equation}
I_{\phi, \text{in}}^{nf}(\delta r_a,\delta r_b)=\frac{4 a \tilde{r}}{Q'(\tilde{r})}I_{r,\text{in}}^{nf}(\delta r_a,\delta r_b)+D(0,\delta r_b),
\end{equation}
\begin{equation}
I_{t,\text{in}}^{nf}(\delta r_a,\delta r_b)=\frac{4 \tilde{r}(a^2+\tilde{r}^2)}{Q'(\tilde{r})}I_{r,\text{in}}^{nf}(\delta r_a,\delta r_b)+H(0,\delta r_b),
\end{equation}
\begin{equation}
I_{\phi, \text{in}}^{fn}(\delta r_a,\delta r_b)=\frac{4 a \tilde{r}}{Q'(\tilde{r})}I_{r,\text{in}}^{fn}(\delta r_a,\delta r_b)+D(\delta r_a,0),
\end{equation}
\begin{equation}
I_{t,\text{in}}^{fn}(\delta r_a,\delta r_b)=\frac{4 \tilde{r}(a^2+\tilde{r}^2)}{Q'(\tilde{r})}I_{r,\text{in}}^{fn}(\delta r_a,\delta r_b)+H(\delta r_a,0).
\end{equation}
Given the definite integral formulas above, we now consider the photon path integral from the source $r_s=\tilde{r}(1+\delta r_s)$ to the observer $r_o=\tilde{r}(1+\delta r_o)$ for photons that reach outside the critical curve:
\begin{equation}
\begin{aligned}
I_r &\approx I^{nf}_{r,\text{out}}(\delta r_0,\delta r_s)+I^{nf}_{r,\text{out}}(\delta r_0,\delta r_o)\\
&=-\frac{1}{2 \tilde{r}\sqrt{\mathcal{\tilde{X}}(\tilde{r})}}\left[ \log\left(\frac{f_1^2-f_2 }{64}\delta r_0^2 \right)+\operatorname{arctanh}\left( \frac{\sqrt{\mathcal{F}(\delta r_s)}}{1+\frac{f_1 \delta r_s}{2}}\right)+\operatorname{arctanh}\left( \frac{\sqrt{\mathcal{F}(\delta r_o)}}{1+\frac{f_1 \delta r_o}{2}}\right)\right] \\
&=-\frac{1}{2 \tilde{r}\sqrt{\mathcal{\tilde{X}}(\tilde{r})}} \log\left[ \frac{(1+\xi_s)(1+\xi_o)}{(1-\xi_s)(1-\xi_o)} \left( \frac{f_1^2-f_2 }{64}\right)\delta r_0^2 \right],
\label{Iout}
\end{aligned}
\end{equation}
where
\begin{equation}
\xi(\delta r)=\frac{\sqrt{\mathcal{F}(\delta r)}}{1+\frac{f_1 \delta r}{2}}, \quad \xi_{s,o,+} \equiv \xi(\delta r_{s,o,+}).
\end{equation}

For the path integral of photons from the source $r_s=\tilde{r}(1+\delta r_s)$ to the event horizon $r_+=\tilde{r}(1+\delta r_+)$ inside the critical curve, we have
\begin{equation}
\begin{aligned}
I_r &\approx I_{r,\text{in}}^{fn}(\delta r_+,\delta r_0)+I_{r,\text{in}}^{nf}(\delta r_0,\delta r_s) \\
&=-\frac{1}{2 \tilde{r}\sqrt{\mathcal{\tilde{X}}(\tilde{r})}}\left[ \log\left(\frac{f_1^2-f_2 }{64}\delta r_0^2 \right)+\operatorname{arctanh}\left( \frac{\sqrt{\mathcal{F}(\delta r_s)}}{1+\frac{f_1 \delta r_s}{2}}\right)+\operatorname{arctanh}\left( \frac{\sqrt{\mathcal{F}(\delta r_+)}}{1+\frac{f_1 \delta r_+}{2}}\right)\right] \\
&=-\frac{1}{2 \tilde{r}\sqrt{\mathcal{\tilde{X}}(\tilde{r})}}\log\left[ \frac{(1+\xi_s)(1+\xi_+)}{(1-\xi_s)(1-\xi_+)} \left( \frac{f_1^2-f_2 }{64}\right)\delta r_0^2\right].
\label{Iin}
\end{aligned}
\end{equation}
The results for the integrals $I_\phi$ and $I_t$ can be obtained in a similar manner.

We now introduce a geometric description of near-critical rays on the observer's screen. In the KBR spacetime, the projection of bound photon orbits onto the screen defines a closed critical curve $\mathcal{C}$ in the image plane $(x,y)$. We parameterize this curve by the radius $\tilde r$ of the corresponding spherical photon orbit:
\begin{equation}
\mathcal{C}:\qquad (x,y) = \bigl(\tilde x(\tilde r),\tilde y(\tilde r)\bigr),
\end{equation}
where the explicit functions $\tilde x(\tilde r)$ and $\tilde y(\tilde r)$ follow from the mapping between conserved quantities and screen coordinates defined in Sec.~\ref{sec:keyparares}. The image-plane metric is Euclidean:
\begin{equation}
ds^2 = dx^2 + dy^2,
\end{equation}
so that the unit tangent and normal vectors to $\mathcal{C}$ are obtained directly from derivatives of $(\tilde x(\tilde r),\tilde y(\tilde r))$. Let $\hat n(\tilde r)$ denote the outward unit normal to $\mathcal{C}$.

Given a point $(x,y)$ in the image plane corresponding to a near-critical ray, we define its signed perpendicular distance $d$ from the critical curve $\mathcal{C}$ by
\begin{equation}
(x,y) - \bigl(\tilde x(\tilde r),\tilde y(\tilde r)\bigr)
= d\hat n(\tilde r),
\qquad d<0 \text{ inside } \mathcal{C}; d>0 \text{ outside } \mathcal{C}.
\label{eq:d-def}
\end{equation}

Near the critical curve, the quantity $\delta r_0^2$ introduced in Eq.~\eqref{delta r_0} can be expressed as a linear function of the screen coordinates:
\begin{equation}
\delta r_0^2 \approx \frac{\partial(\delta r_0^2)}{\partial x} \bigl[x-\tilde x(\tilde r)\bigr] + \frac{\partial(\delta r_0^2)}{\partial y} \bigl[y-\tilde y(\tilde r)\bigr].
\end{equation}
Its gradient is proportional to the outward normal:
\begin{equation}
\vec\nabla(\delta r_0^2)
= \bigl|\vec\nabla(\delta r_0^2)\bigr|\hat n(\tilde r).
\end{equation}

Restricting to displacements perpendicular to $\mathcal{C}$, we obtain
\begin{equation}
\delta r_0^2 = K(\tilde r) d, \qquad K(\tilde r) = \bigl|\vec\nabla(\delta r_0^2)\bigr| = \sqrt{\left( \frac{\partial (\delta r_0^2)}{\partial x} \right)^2 + \left( \frac{\partial (\delta r_0^2)}{\partial y} \right)^2},
\label{dr0-d}
\end{equation}
where $K(\tilde r)$ is a positive function of the bound-orbit radius. Substituting Eq.~\eqref{dr0-d} into Eq.~\eqref{Iout} and Eq.~\eqref{Iin}, and absorbing the finite contributions into a single constant inside the logarithm, we may finally write
\begin{equation}
I_r \approx -\frac{1}{2\tilde r \sqrt{\tilde X(\tilde r)}} \log\left[ C_\pm d \right],
\end{equation}
where
\begin{equation}
C_+ \equiv C(\tilde{r};r_s,r_o) = \frac{(1+\xi_s)(1+\xi_o)}{(1-\xi_s)(1-\xi_o)} \left( \frac{f_1^2-f_2}{64} \right) K(\tilde r),
\end{equation}
\begin{equation}
C_- \equiv C(\tilde{r};r_s,r_+) = -\frac{(1+\xi_s)(1+\xi_+)}{(1-\xi_s)(1-\xi_+)} \left( \frac{f_1^2-f_2}{64} \right) K(\tilde r).
\end{equation}
$C_\pm$ is a finite function that encodes the details of the near- and far-region contributions.

Following the same approach, the results for $I_\phi$ and $I_t$ can be written as
\begin{equation}
I_\phi \approx \frac{4 a \tilde r}{Q'(\tilde r)} I_r + D_\pm,
\end{equation}
\begin{equation}
I_t \approx \frac{4 \tilde r (a^2 + \tilde r^2)}{Q'(\tilde r)} I_r + H_\pm,
\end{equation}
where the finite remainders $D_\pm$ and $H_\pm$ are
\begin{equation}
D_+ \equiv D(\tilde r; r_s, r_o) = D(0,\delta r_s) + D(0,\delta r_o),
\end{equation}
\begin{equation}
D_- \equiv D(\tilde r; r_s, r_+) = D(0,\delta r_s) + D(\delta r_+,0),
\end{equation}
\begin{equation}
H_+ \equiv H(\tilde r; r_s, r_o) = H(0,\delta r_s) + H(0,\delta r_o),
\end{equation}
\begin{equation}
H_- \equiv H(\tilde r; r_s, r_+) = H(0,\delta r_s) + H(\delta r_+,0).
\end{equation}

Notably, the subscript $\pm$ uniformly denotes the image point's location relative to the inner/outer side of the critical curve $\mathcal{C}$, i.e., $\pm = \operatorname{sign}(d)$ (where $d>0$ corresponds to “outside” and $d<0$ corresponds to “inside”). Therefore, $C_\pm$, $D_\pm$, and $H_\pm$ take two sets of constants depending only on the sign of $d$, independent of the magnitude of $|d|$.

\bibliographystyle{utphys}
\bibliography{KBRlensing}


\end{document}